# Contact-inhibited chemotaxis in *de novo* and sprouting blood-vessel growth

Roeland M. H. Merks [1,2,3,*], Erica D. Perryn [4,+], Abbas Shirinifard [3], and James A. Glazier [3]

[1]VIB Department of Plant Systems Biology, Technologiepark 927, B-9052 Ghent, Belgium

[2]Department of Molecular Genetics, Ghent University, B-9052 Ghent, Belgium

[3]The Biocomplexity Institute and Department of Physics, Indiana University Bloomington, Swain Hall West 159, 727 East 3rd Street, Bloomington, IN 47405-7105, USA

[4]The University of Kansas Medical Center, Department of Anatomy and Cell Biology, 1008 Wahl Hall West, 3901 Rainbow Boulevard, Kansas City, KS 66160, USA

[+] Present address: Krumlauf Laboratory, Stowers Institute for Medical Research, 1000 East 50th Street, Kansas City, MO 64110, USA

[*] **Corresponding author:** post@roelandmerks.nl, roeland.merks@psb.ugent.be

**E-mail:** `post@roelandmerks.nl`, `EDP@stowers-institute.org`, `ashirini@indiana.edu, glazier@indiana.edu`

Contact-inhibited chemotaxis in *de novo* and sprouting blood vessel growth

# Abstract

**Background**

Blood vessels form either when dispersed *endothelial* cells (the cells lining the inner walls of fully-formed blood vessels) organize into a vessel network (*vasculogenesis*), or by sprouting or splitting of existing blood vessels (*angiogenesis*). Although they are closely related biologically, no current model explains both phenomena with a single biophysical mechanism. Most computational models describe sprouting at the level of the blood vessel, ignoring how cell behavior drives branch splitting during sprouting.

**Methodology/principal findings**

We present a cell-based, Glazier-Graner-Hogeweg-model simulation of the initial patterning before the vascular cords form lumens, based on plausible behaviors of endothelial cells. The endothelial cells secrete a chemoattractant, which attracts other endothelial cells. As in the classic Keller-Segel model, chemotaxis by itself causes cells to aggregate into isolated clusters. However, including experimentally-observed adhesion-driven contact inhibition of chemotaxis in the simulation causes randomly-distributed cells to organize into networks and cell aggregates to sprout, reproducing aspects of both *de novo* and sprouting blood-vessel growth. We discuss two branching instabilities responsible for our results. Cells at the surfaces of cell clusters attempting to migrate to the centers of the clusters produce a buckling instability. In a model variant that eliminates the surface-normal force, a dissipative mechanism drives sprouting, with the secreted chemical acting both as a chemoattractant and as an inhibitor of pseudopod extension. Both mechanisms would also apply if force transmission through the extracellular matrix rather than chemical signaling mediated cell-cell interactions.

**Conclusions/significance**

Contact-inhibited chemotaxis in *de novo* and sprouting blood vessel growth

The branching instabilities responsible for our results, which result from contact inhibition of chemotaxis, are both generic developmental mechanisms and interesting examples of unusual patterning instabilities.

Contact-inhibited chemotaxis in *de novo* and sprouting blood vessel growth

# Synopsis

During embryonic development, *endothelial* cells (the cells lining the inner walls of blood vessels) initially self-organize into a network of solid cords via *vasculogenesis*. This *primary vascular plexus* then develops into capillaries as the cords hollow out (form *lumens*) by a variety of mechanisms. Further remodeling and association with additional cell types leads finally to the mature vascular network. In adults, the vascular network continues to expand by splitting of existing blood vessels and by sprouting. Such *angiogenesis* is crucial during tumor growth, diabetes-related conditions, and wound healing. Using a computer simulation, we have captured a small set of biologically-plausible cell behaviors that can reproduce the initial self-organization of endothelial cells, the sprouting of existing vessels and the immediately subsequent remodeling of the resulting networks. Endothelial cells both secrete diffusible *chemoattractants* and move up gradients of those chemicals by extending and retracting small *pseudopods*. By itself, such *chemotaxis* causes simulated cells to aggregate into large, round clusters. However, endothelial cells stop extending pseudopods along a given section of cell membrane as soon as the membrane touches the membrane of another endothelial cell (*contact inhibition*). Adding contact-inhibition to our simulations allows vascular cords to form sprouts under a wide range of conditions. We conclude by testing experimentally the simulation's prediction that network development requires contact-inhibited chemotaxis.

Contact-inhibited chemotaxis in *de novo* and sprouting blood vessel growth

# 1. Introduction

*1.1 Vasculogenesis and Angiogenesis*

Blood-vessel development is essential for myriad biological phenomena in healthy and diseased individuals, including wound healing and tumor growth [1,2]. Blood vessels form either *de novo*, via *vasculogenesis* or by sprouting or splitting of existing blood vessels via *angiogenesis*.

In vasculogenesis, dispersed *endothelial cells* (*ECs*; the cells lining the inner walls of fully-formed blood vessels) organize into a primary vascular plexus of solid cords which then remodel into a vascular network. ECs elongate parallel to the cords, with final aspect ratios of tens to one. Because the early stages of vasculogenesis depend on a single cell type, vasculogenesis is relatively easy to reproduce *in vitro*. When cultured *in vitro* on *Matrigel*, a commercial product mimicking the extracellular matrix (*ECM*; the mixture of proteins, growth-factors and carbohydrates surrounding cells *in vivo*), even in the absence of other cell types or positional cues, ECs organize into cords which form large-scale, honey-comb like patterns, with cords of ECs surrounding regions devoid of ECs. This network slowly reorganizes, with the size of the polygonal, cell-free *lacunae*, gradually increasing. This observation suggests that ECs have *autonomous* patterning ability, rather than following morphogen pre-patterns.

The sprouting or splitting of existing blood vessels during *angiogenesis* is more complex. In the first step of angiogenesis, a vessel dilates and releases plasma proteins that induce a series of changes in EC behavior. The ECs which will form the sprout next detach from each other and from the surrounding smooth-muscle cells, destabilizing the vessel. These detached ECs proliferate, migrate out of the vessel and organize into a sprout. EC proliferation continues in the sprout and is fastest just behind the leading *tip cell*, which is selected using a lateral-inhibition mechanism mediated by Dll4 and Notch1 [3]. Finally, the sprout forms a lumen, secretes a basal lamina and associates with pericytes that stabilize the sprout to form a mature new vessel [4].

Contact-inhibited chemotaxis in *de novo* and sprouting blood vessel growth

Two fundamental questions concerning vasculogenesis and angiogenesis and their relation to each other are: 1) Does blood-vessel formation require external patterning cues (*pre-patterns* of morphogens) to define the precise position of the ECs, or can ECs organize into vascular patterns autonomously, with external cues merely initiating and fine-tuning vascular morphogenesis? 2) Do vasculogenesis and angiogenesis require the same or different cell behaviors, molecular signals and biomechanics?

*1.2 Experimental Background*

Despite the biomedical importance of angiogenesis and vasculogenesis, existing experiments are sufficiently ambiguous that even the fundamental mechanisms guiding patterning are uncertain. Experiments suggest a central role for chemotaxis in both *de novo* and sprouting blood-vessel growth [4-6]. ECs respond to, and often produce, a wide range of chemoattractants and chemorepellants, including the many isoforms of vascular-endothelial growth factor A (*VEGF-A*) [6], the chemokine SDF-1 [7,8], which ECs secrete [7], fibroblast growth factor 2 (*FGF-2*), which induces ECs in developing vessels to secrete VEGF [9], Slit-2, which can act either as a chemoattractant or a chemorepellant depending on the receptor to which it binds [10], and the chemorepelling semaphorins [10].

Which of these molecules (if any) govern vascular patterning is still unclear. The Torino Group (*e.g.* [11,12]) argued that a VEGF-A was the short-range autocrine chemoattractant that their chemotaxis-based blood-vessel-growth model required, since ECs express receptors for VEGF (*VEGFR-2*), chemotax towards sources of VEGF under favorable conditions, and secrete VEGFs. However, experiments suggest that cell-autonomous secretion of VEGF is essential only for vascular maintenance, not for angiogenesis *per se*: mice genetically-engineered to lack the VEGF gene only in their ECs have normal vascular density and patterning, but impaired vascular homeostasis and EC survival [13]. A plausible, alternative cell-autonomous chemoattractant to

Contact-inhibited chemotaxis in *de novo* and sprouting blood vessel growth guide EC aggregation is the chemokine SDF-1/CXCL12, which ECs both secrete and respond to [8].

However, based on experiments that suggest that ECs can follow stresses in the ECM (see *e.g.* [14] for review), Manoussaki and Murray [15], and Namy *et al.* [16] proposed that mechanical interactions rather than, or in addition to, chemical interactions govern vasculogenesis. Further complicating this picture, Szabo and coworkers [17] showed that non-vascular, glia or muscle cells cultured on rigid, plastic culture dishes in continuously-shaken medium can form linear structures. Such culture conditions should reduce both the formation of chemoattractant gradients or migration along stress lines in the ECM. In the absence of ECM, they hypothesized that cells preferentially move towards elongated structures. Szabo and coworkers [17] proposed two mechanisms for such cell behavior: cells would align to surrounding cells, or they would mechanotactically follow stress fields in the cytoskeleton of neighboring cells. However, the molecular mechanisms of such cell behavior remains unclear as is the relevance of these results to ECs.

Angiogenesis and vasculogenesis also require a number of local, contact-dependent (*juxtracrine*) signals: Tip-cell selection during angiogenic sprouting depends on Delta-notch signaling [3], while Eph receptor-ephrin ligand binding amplifies ECs' response to SDF-1 [8]. All ECs express vascular-endothelial-cadherin (*VE-cadherin*), a homophilic, trans-membrane cell-adhesion molecule, which appears to play a crucial role in vascular patterning [18,19]. Besides its role in cell-cell adhesion, VE-cadherin has a signaling function that determines how ECs respond to VEGF-A. When ECs bind to other ECs through their VE-cadherin, VEGF-A reduces their motility and proliferation. In the absence of VE-cadherin binding, VEGF-A activates pathways related to actin polymerization and the cell cycle, enhancing cell motility and proliferation in sub-confluent monolayers, and causes preferential extension of pseudopods in directions with higher VEGF-A concentrations [20]. We hypothesize that VE-cadherin-binding acts locally to prevent extension of pseudopods in the direction of cell-cell contacts for all critical chemoattractants, not only to VEGF-A. VE-cadherin -/- double-knock-out mice develop abnormal vascular networks in the yolk sac

Contact-inhibited chemotaxis in *de novo* and sprouting blood vessel growth

[18], with ECs forming isolated vascular islands instead of wild-type polygonal vascular networks. These mice also have defective angiogenic sprouting, suggesting that both vasculogenesis and angiogenesis require VE-cadherin. VE cadherin -/- ECs still form strong adhesive junctions, so loss of VE-cadherin-mediated signaling rather than loss of intercellular adhesion seems to be responsible for the knock-out phenotype [18].

*1.3 Computational Background*

Fortunately, a number of models and simulations replicate features of *in vitro* vascular patterning and can help partially reconstruct minimal sets of behaviors ECs require to self-organize into polygonal, vascular patterns [11,12,15-17,21-23].

Because of the experiments we discussed above and others which have demonstrated that sprouting angiogenesis and vasculogenesis both require chemotaxis (see, *e.g.*, [7,8,24]), most models of vasculogenesis assume that intercellular signaling occurs via a diffusible chemoattractant. Using continuum models deriving from the fluid-dynamic Burgers' equation, Preziosi and coworkers (called the *Torino Group* in this paper) showed that simulated ECs secreting a chemoattractant that attracts surrounding ECs, could self-organize into polygonal patterns similar to the patterns in EC cultures and *in vivo* [11,12,25,26]. However, their work assumed that endothelial cells accelerate in chemical gradients, which is not plausible in the highly viscous, non-inertial environment of the ECM. Microfluidic evidence indicates that mammalian cells (HL60) rapidly reach a flow-dependent, constant velocity [27] in chemoattractant gradients rather than continuously accelerating. We have previously suggested that [22] a linear force-velocity relation is the most appropriate model of ECs' experimental response, with the velocity of ECs proportional to the strength of the gradient of the chemoattractant. However, in simulations of this simple model, isotropic ECs form well-separated rounded clusters instead of networks. We have shown that adding one of a number of mechanisms (including cell adhesion [21] and cell elongation [22]) to

Contact-inhibited chemotaxis in *de novo* and sprouting blood vessel growth chemotactic aggregation suffices to produce quasi-polygonal networks. Section 2 discusses these mechanisms in more detail.

In the mechanical models of Manoussaki and Murray [15], and Namy *et al.* [16] ECs pull on the elastic ECM and aggregate by *haptotactically* migrating along the resulting ECM stress lines. Surprisingly, the mathematical form of the chemical and mechanical models is practically identical. Because these mechanical models assume that ECs exert radially-symmetric stresses on the ECM, modeling stress fields and EC haptotaxis or EC secretion and response to a chemoattractant, results in the same cell movement. Since simulations of the two mechanisms are identical, distinguishing between the effects of chemical and mechanical mechanisms will require additional experiments (such experiments are currently underway in the Glazier laboratory (Shirinifard, Alileche and Glazier, *preprint, 2008*)).

A separate set of simulations address angiogenesis. Many models of sprouting blood-vessel growth introduce blood-vessel-level phenomenology by hand through high-level rules for branching [28-30]. Attempts to derive blood-vessel sprouting and splitting from the underlying behavior of ECs include Levine and coworkers' [31] model of the onset of angiogenic sprouting as a reinforced random walk, where the ECs degrade the ECM, which locally enhances EC motility and produces paths of degraded ECM, and Bauer and Jiang's [32] cell-based model of blood-vessel sprouting along externally-generated morphogen gradients, which assumed that branch splitting results from ECM inhomogeneities. Neither model can explain both EC assembly and blood-vessel sprouting.

Could the behavior of the individual ECs also explain aspects of blood-vessel sprouting? Because the same genetic machinery regulates both angiogenesis and vasculogenesis [4], a common set of mechanisms is plausible. Manoussaki [33] extended her mechanical model of vasculogenesis to describe angiogenesis by adding long-range, chemotactic guidance cues. In her simulations, ECs migrated from an aggregate towards a chemoattractant source and cell-traction-driven migration contracted the sprout into a narrow, vessel-like cord.

Contact-inhibited chemotaxis in *de novo* and sprouting blood vessel growth

In this paper we present an alternative chemotaxis-based mechanism that can produce networks both from dispersed ECs and EC clusters without requiring long-range guidance cues. Instead, in our model long-range signals would only steer the self-organized vessels, a more biologically-realistic mechanism. Extending simulations that we have briefly introduced elsewhere [23], we show that VE-cadherin-mediated contact inhibition of chemotactic pseudopod projections, in combination with secretion of a diffusing, rapidly decaying chemoattractant by ECs, suffices to reproduce aspects of both *de novo* and sprouting blood-vessel growth. In our simulations ECs: a) secrete a chemoattractant and b) preferentially extend pseudopods up gradients of the chemoattractant, unless, c) contact inhibition locally prevents chemotactic pseudopod extension. Thus, cell-cell binding suppresses the extension of chemotactic pseudopods, while unbound cell surfaces in contact with the ECM continue to extend pseudopods towards sources of chemoattractant [24]. We compare two biologically-plausible scenarios for chemotaxis, one in which ECs actively extend and retract pseudopods along chemoattractant gradients, and one in which the pseudopods' retractions are chemotactically neutral. The second scenario suggests a sprouting mechanism where a secreted autocrine factor acts both as a long-range chemoattractant and a local inhibitor of pseudopod sprouting.

## 2. Results

We modeled endothelial cell behavior at a mesoscopic level using the Glazier-Graner-Hogeweg (*GGH*) model, also known as the Cellular Potts Model (*CPM*) [34-37]. The GGH is a lattice-based Monte-Carlo approach that describes biological cells as spatially-extended patches of identical lattice indices. Intercellular junctions and cell junctions to the ECM determine adhesive (or binding) energies. The GGH algorithm, which we describe in more detail in Section 4 (Materials & Methods), models pseudopod protrusions by iteratively displacing cell interfaces, with a preference for displacements which reduce the local *effective energy* of the configuration. Cells reorganize to favor stronger rather than weaker cell-cell and cell-ECM bonds and shorter rather than longer cell

Contact-inhibited chemotaxis in *de novo* and sprouting blood vessel growth boundaries. In addition to interface displacements that reduce the effective energy, active cell motility also allows displacements that increase the effective energy. The likelihood of these active displacements increases with the *cell-motility* parameter $T$. Further constraints regulate cell volumes, surface areas, and chemotaxis. To model chemotaxis, we use the Savill and Hogeweg [36] algorithm that favors extensions of pseudopods up concentration gradients of a chemoattractant (see Eq. 3 in Section 4). In the simplest implementation of chemotaxis in the GGH, cell velocity is proportional to the strength of the chemical gradient, in general agreement with experiments; see *e.g.* [22] (we discuss the details of chemotaxis implementation below in the subsections **2.5** and **2.6** and in Section 4; see especially Eq. 3).

The advantage of the GGH over alternative cell-based modeling approaches [38] that represent cells as point particles or fixed-sized spheres or ellipsoids is that we can differentiate between bound and unbound regions of cell membrane. The GGH naturally represents the stochastic, exploratory behavior of migrating cells, modeling it as the biased extension and retraction of pseudopods, instead of a biologically-implausible single force acting on cells' centers of mass as in some cell-based simulations.

We described chemoattractant diffusion and degradation macroscopically, using a continuum approximation. In analogy to the Torino Group's continuum model of *de novo* blood-vessel growth [12,25], ECs secrete a diffusing chemoattractant at a rate $\alpha$, which degrades in the ECM at a rate $\varepsilon$ (*e.g.* due to proteolytic enzymes or by binding to ECM components), obeying:

$$\frac{\partial c}{\partial t} = \alpha(1 - \delta(\sigma(\vec{x}),0)) - \varepsilon\delta(\sigma(\vec{x}),0)c + D\nabla^2 c, \qquad (1)$$

where $\delta(\sigma(\vec{x}),0) = 0$ inside cells and is $\delta(\sigma(\vec{x}),0) = 1$ in the ECM. Because we wish to compare our simulations to experimental yolk-sac cultures, where the vascular patterns are essentially monolayers, we use a two-dimensional GGH.

Contact-inhibited chemotaxis in *de novo* and sprouting blood vessel growth

We set the chemoattractant's secretion rate by cells $\alpha = 10^{-3} \text{s}^{-1}$, its decay rate $\varepsilon = \alpha$, and its diffusion constant in ECM to a slow $D = 10^{-13}$ m$^2$s$^{-1}$. These parameter values produce steeper gradients than those for VEGF-A$_{165}$, the chemoattractant which Gamba *et al.* suggested was responsible for vasculogenesis, which has a diffusion coefficient of $D \sim 10^{-11}$ m$^2$s$^{-1}$ [12]. The diffusion coefficient of SDF-1/CXCL12 is in the range of $1.7 \times 10^{-13}$ m$^2$s$^{-1}$ [39]. However, the phenomena we observe in our simulations hold over a large range of diffusion coefficients.

*2.1 EC aggregation and vasculogenesis in the absence of contact inhibition*

In Fig. 1(A-C) and Movie S1, we randomly distributed 1000 ECs, each with an area of $\sim 200\,\mu\text{m}^2$ over an area of $\approx 700\,\mu\text{m} \times 700\,\mu\text{m}$ ($333 \times 333$ lattice sites, or *pixels*, of $2\,\mu\text{m} \times 2\,\mu\text{m}$ each), which we positioned inside a larger lattice of $1000\,\mu\text{m} \times 1000\,\mu\text{m}$ to minimize boundary effects. In this cell-based simulation of the Torino Group's continuum model [11,12], without endothelial-cell acceleration in chemoattractant gradients our cells form disconnected, vascular islands rather than a vascular network. We would expect this result, because, with the more realistic chemotactic response we employ, the Torino Group's model reduces to the classic Keller-Segel equations [40] of chemotactic aggregation [25], which, like our simulations, form isolated vascular islands. Apparently, the basic Torino-Group model of chemotactic cell aggregation misses a biological mechanism essential for vasculogenesis. We have previously suggested a number of additional mechanisms, any one of which, together with cell aggregation, suffices to induce vasculogenesis-like patterning. *E.g.,* when we gave the ECs the elongated shapes observed in later stages of experiments, neighboring cells aligned with each other, causing cell clusters to elongate and interconnect, creating a vascular network, in a mechanism similar to Szabo's [17]. These vascular networks remodel gradually, with dynamics resembling those of *in-vitro* vascular networks. The causes of cell elongation in experiments are not clear. ECs could elongate either cell-autonomously (*e.g.* by remodeling their cytoskeletons), or non-cell-autonomously, by maximizing their contact

Contact-inhibited chemotaxis in *de novo* and sprouting blood vessel growth areas with surrounding cells or by aligning to morphogen gradients in the ECM [22]. Unless we state otherwise, in this paper we neglect cell-autonomous elongation.

Even without strong cell-cell adhesion the ECs can form vascular-like structures in simulations of vasculogenesis if the *diffusion length* of the chemoattractant (the length $L$ over which the concentration drops to half its value at the EC membrane) is short enough, because the ECs align with the chemical gradients [23]. This length scale $L$ depends on the diffusion coefficient $D$ and the chemoattractant decay rate $\varepsilon$ as $L = \sqrt{\frac{D}{\varepsilon}}$ [12].

## 2.2 Sprouting angiogenesis in the absence of contact-inhibition

To investigate whether the Torino-Group Model could reproduce sprouting angiogenesis, we started our simulations with rounded clusters of simulated ECs representing a blood vessel's surface after degradation of the ECM, keeping the simulation parameters unchanged from Fig. 1. As we expected, the clusters of ECs did not form sprouts, Fig. 4(C).

As in vasculogenesis, cell-elongation sufficed to drive angiogenesis-like sprouting (see Fig. 2(G-I)), where we used a length constraint, see [22]). EC clusters also produced sprouts for strong cell-cell adhesion (*i.e.* for values of $J(c,c)<10$) (Fig. 2(A-C)), via a mechanism similar to the cell-elongation-dependent mechanism for vasculogenesis [22]. Adhesion-independent sprouting occurred only for a narrow range of very small diffusion constants of the chemoattractant, between $D < 2 \cdot 10^{-14}$ m$^2$s$^{-1}$ and $D > 4 \cdot 10^{-14}$ m$^2$s$^{-1}$ (see Fig. 2(D-F)). The allowable range of $D$ increased for bigger cells [23].

Figs. 7-10 and supplementary Fig. S1 show the results of systematic screens for sprouting in the absence of contact-inhibited chemotaxis, but we defer an in-depth study of these phenomena to our future work.

*2.3 Contact-inhibited chemotaxis in* de novo *blood vessel growth*

In this paper, we focus on the role of contact-inhibited chemotaxis in sprouting blood-vessel growth. We hypothesize that VE-cadherin's local inhibition of chemotaxis-induced pseudopod extensions at EC-EC boundaries, may be responsible for ECs' self-organization into vascular-like networks.

We modeled contact inhibition of chemotaxis in our simulations by suppressing chemotaxis at cell-cell interfaces. Thus, only interfaces between cells and ECM respond to the chemoattractant. Fig. 1(D) and Supplementary Movies S2 and S3 show typical simulations of *de novo* blood-vessel growth with contact inhibition. The ECs assemble into a structure resembling a capillary plexus: cords of cells enclose lacunae, which grow slowly. Smaller lacunae shrink and disappear, while larger lacunae subdivide via vessel sprouting as, *e.g.*, in the quail yolk sac [41].

*2.4 Contact-inhibited chemotaxis in blood vessel sprouting*

To investigate the role of contact-inhibited chemotaxis in blood vessel sprouting, we ran a set of simulations with a large cluster of endothelial cells representing a blood vessel's surface after degradation of the ECM, keeping all simulation parameters the same as those in

The surface of the cluster first roughens, with some cells protruding from the surface, then digitates into a structure reminiscent of a primary vascular plexus (Fig. 4 (A-C) and Movies S4 and S5), the first type of structure to develop in both *de novo* and sprouting blood-vessel growth [41]. The sprouting instability requires contact inhibition of chemotaxis. Without it, the clusters remained rounded and compact (Fig. 4(D)). Thus our simulations suggest that a process operating at the level of individual cells—chemotaxis with contact inhibition—may drive *in vitro* blood-vessel growth both sprouting and *de novo*.

What drives blood vessel sprouting in our model? At equilibrium, the chemoattractant has a quasi-Gaussian profile across the cluster. It levels off towards the cluster's center, while its

Contact-inhibited chemotaxis in *de novo* and sprouting blood vessel growth inflection point is at the cluster boundary. Chemotaxis produces a continuous, inward, normal force at the cluster boundary, creating a buckling instability (see *e.g.* [42]); chemotactic forces also compress small initial bumps laterally, producing sprouts. Since contact inhibition of chemotaxis leaves the interior cells insensitive to the chemoattractant, ingressing surface cells easily push them aside. When we omit contact inhibition of motility to mimic anti-VE-cadherin-antibody-treated allantois cultures, the interior cells also feel the inward-directed chemotactic forces and resist displacement (Fig. 4 (D) and Movie S6).

To explore this idea, we varied the ratio of the chemotactic response at cell-cell interfaces relative to the chemotactic response at cell-ECM interfaces ($\chi(c,c)/\chi(c,M)$), where $\chi(c,c)$ is the ECs' sensitivity to the chemoattractant at cell-cell interfaces and $\chi(c,M)$ the sensitivity at cell-ECM interfaces (see Section 4 for details). We looked for sprouting in clusters of 128 cells, each of area $\sim 200\,\mu m^2$, placed in a $400\,\mu m \times 400\,\mu m$ lattice, keeping all other parameters unchanged from their values in Fig. 4.

We defined the clusters' compactness after 10000 Monte Carlo Steps (the time unit of the simulation, see Section 4, with 1 MCS equivalent to about 30 s) to be $C = A_{cluster}/A_{hull}$, the ratio between the cluster's area, $A_{cluster}$, and the area of its convex hull (that is the tightest possible "gift wrapping" around the cluster), $A_{hull}$. The compactness $C = 1$ for a perfectly circular cluster, whereas $C \to 0$ for highly branched or dispersed clusters of cells.

We found a phase transition at $\chi(c,c)/\chi(c,M) \approx 0.5$ separating sprouting from non-sprouting clusters (Fig. 5), suggesting that the sprouting instability only occurs when the core of the cluster behaves as a fluid: because each cell's volume is nearly conserved (apart from small fluctuations around its target volume), the core cells can only release the pressure the ingressing cells exert on them by moving outwards as sprouts. Our ongoing work characterizes this instability mathematically, proving that the cluster self-organizes into a network structure with fixed cord width (A. Shirinifard and J. A. Glazier, *preprint 2008*).

Contact-inhibited chemotaxis in *de novo* and sprouting blood vessel growth

To validate our model against published EC tracking experiments [19], we compared the trajectories of cells in sprouting and non-sprouting clusters. Fig. 6(A-D) show the trajectories of ten cells in a sprouting cluster (with contact-inhibition; panels a-b), and ten cells in a non-sprouting cluster (without contact-inhibition; panels c-d). In non-sprouting clusters, cells followed random-walk trajectories, while in sprouting clusters, they followed biased random-walk trajectories. To further characterize cell motility, we measured cells' average displacements and velocities over 10 independent simulations of 128 cells each. In sprouting clusters, the cells moved further during a given interval than in non-sprouting clusters. Thus, the cell velocity $V_i(t) = (\vec{x}_i(t+\Delta t) - \vec{x}_i(t-\Delta t))/2\Delta t$ [19] is larger during sprouting if the interval $\Delta t$ between subsequent cell positions is sufficiently large (here we use $\Delta t = 2.5$ h as in Perryn *et al.* [19]); for shorter intervals (*e.g.* 30 s) the cell velocity is highest in non-sprouting clusters (not shown), indicating that ECs in sprouting clusters moved faster, but had a somewhat slower random motility.

Our simulations agree with recent experiments tracking ECs in embryonic mouse allantoides [19] that measured the cell-autonomous motility of ECs cells in allantoides relative to the motility of the surrounding mesothelium in which the ECs reside. Administration of anti-VE-cadherin antibodies reduced both cell-autonomous motion and net displacement of ECs. Thus, our simulations suggest that VE-cadherin's role as a contact-dependent inhibitor of cell motility suffices to explain the reduced cell motility observed in anti-VE-cadherin-treated allantoides cultures.

*2.5 Sensitivity analysis*

Contact-inhibited sprouting occurs for a wide range of parameter values. In most of our simulations we set the EC-EC adhesion equal to the EC-ECM adhesion (*i.e.* $J(c,c) = 2J(c,M)$; the factor of 2 arises because we model the ECM as a single large generalized cell), which is equivalent to setting the surface tension of the cluster to zero [35]. Zero surface tension clarifies the role of contact inhibition in sprouting, but real ECs adhere strongly to each other via *adherens junctions* [18]. In Fig. 7 and in Movies S7A-P we studied the effect of cell-cell adhesion on sprouting in clusters of

Contact-inhibited chemotaxis in *de novo* and sprouting blood vessel growth

128 cells (256 cells in the movies). For stronger EC-EC adhesion, equivalent to positive surface tension, $J(c,c) < 2J(c,M)$, the sprouts are longer and thinner and the network less compact than for zero surface tension. For very weak EC-EC adhesion ($J(c,c) \gg 2J(c,M)$), equivalent to strong negative surface tension, the ECs separate from each other, so contact-inhibition no longer occurs, and the clusters do not sprout. For small negative surface tensions, with values of $J(c,c) > 2J(c,M)$, chemotaxis overcomes the negative surface tension, so ECs still touch each other and sprouting occurs as for zero surface tension, producing thickened sprouts and elongated clusters. The insets to Fig. 7 and Movies S7N to S7P show the results for $50 \leq J(c,c) \leq 70$.

We also investigated how sprouting depends on the chemotactic strength $\chi(c,M)$ (Fig. 8 and Movies S8A-K). For $\chi(c,M) = 500$, most vascular cords are two cells wide (Movie S8B), while for $\chi(c,M) > 500$ the cords become thinner and longer, with cords only one cell wide (Movies S8C-K). For higher chemotactic forces, the cells intercalate, moving to the chemical gradients' peak. We have derived the conditions for this *folding instability* in our ongoing work (A. Shirinifard and J. A. Glazier, *preprint, 2008*). Higher chemotactic strengths increase ruffling of the cluster boundary, reducing the cluster's compactness in the absence of contact inhibition (Fig. 8).

We assumed that ECs extend or retract pseudopods depending on the difference in chemoattractant concentration between the retracted and extended positions, independent of the absolute chemoattractant concentrations. However, at higher chemoattractant concentrations, most chemoattractant receptors will saturate with chemoattractant and become insensitive to chemoattractant levels. To study the effect of saturated chemotactic response [21] on angiogenic sprouting, we varied the saturation parameter *s* (see Eq. 3, Section 4) leaving all other parameters unchanged. For $s = 0$, the chemotactic response is linear; for $s > 0$, the response to the chemoattractant gradient vanishes at high concentrations (see Section 4). For small positive *s*, the clusters sprout normally (see Fig. 9 and Movies S9A-C); however, for large *s*, the chemotactic response weakens at the chemoattractant levels present at the edge of the cell cluster; thus cells no longer chemotact towards the cluster's interior and the sprouting instability disappears (Movies

Contact-inhibited chemotaxis in *de novo* and sprouting blood vessel growth

S9D-E). We could test this prediction experimentally by partially inactivating the ECs' chemoattractant receptors. We observed the same effect when we increased the chemoattractant secretion rate for moderate response saturation ($s$=0.05; see Supplementary Fig. S2, bottom panel) leading to higher overall chemoattractant concentrations. We could test this situation experimentally by overexpressing the chemoattractant in ECs. Since for unsaturated chemotactic response ($s$=0), multiplying the chemoattractant concentrations is equivalent to multiplying the chemotactic strength ($\chi(c,M)$) by the same factor, increasing the secretion rate first thins and lengthens the cords by increasing the chemotactic strength, then eventually prevents sprouting as the chemotactic response saturates. This effect is most apparent for $s = 0.01$ (Fig. S2, top panel).

In the Torino Group's continuum model, the separation between the cords increases with the diffusion length $L$ of the chemoattractant, Fig. 10 and Movies S10A-G show sprouting clusters for a range of diffusion lengths. In agreement with the Torino Group's model, longer diffusion lengths produce thicker cords with larger intercord spaces. The clusters do not sprout well when $L$ approaches the EC-cluster diameter. Clusters consisting of 1024 cells sprout for $D > 3 \cdot 10^{-13}$ m$^2$s$^{-1}$ ($L > 17.3$ μm), while 128-cell clusters do not (Fig. 10 and Movies S11A-G). If the diffusion length is smaller than the ECs' diameter, the clusters dissociate: the ECs perform random walks with long persistence lengths, moving up the chemoattractant gradients they leave behind themselves (Movies S10A and S11A).

## *2.6 A dissipative sprouting mechanism*

In our simulations, the trailing edges of the ECs retract actively in response to the chemoattractant and exert an inward-normal, compressive force on the EC cluster. To check if sprouting requires this compressive force, we also simulated a situation in which only *extending* pseudopods at cell-ECM interfaces respond to the chemoattractant, while retraction is chemotactically neutral. Both sprouting-angiogenesis and vasculogenesis occurred, but required higher intrinsic cell motilities (larger values of the parameter $T$). Fig. 11 shows the motilities required under both assumptions.

Contact-inhibited chemotaxis in *de novo* and sprouting blood vessel growth

We looked for sprouting after 5000 MCS (~ 40 h) in clusters of 128 cells, each of area $\approx 200\ \mu m^2$, placed in a $400\ \mu m \times 400\ \mu m$ lattice, with all other parameters the same as in Fig. 4. For $T < 100$, our original chemotaxis assumptions produced sprouts, while no sprouting occurred if pseudopods responded to the chemoattractant only during extension. For $100 < T < 400$, both mechanisms produced sprouts. For $T > 400$, the ECs broke up into small pieces, a well-characterized, non-biological artifact of the GGH [35]. With extension-only chemotaxis, sprouting was slightly slower than for standard, extension-retraction Savill-Hogeweg [36] chemotaxis, as a plot of the time evolution of the clusters' compactness shows (Fig. 12 and Movies S12-S14). However, at long times ($t > 2500\ MCS$) the compactness of clusters decreased at identical rates for both methods.

These results suggest an additional mechanism for blood-vessel sprouting: at the cluster surface, *all* pseudopod extensions increase the effective energy slightly, so the chemoattractant *inhibits* pseudopod extension. A recent experimental study [43] found that autocrine secretion of the sprouting inhibitor TGF-β1 enhances branching in mammary epithelial tubes. Our model suggests a mechanism by which an autocrine, secreted chemical can act *both* as a chemoattractant and as an inhibitor. The rates of pseudopod extensions and retractions are critical to pattern evolution (Fig. 11). Cells in growing tips see a shallower gradient than do those in valleys between the tips (see *e.g.* Fig. 4(B)), so pseudopod extensions at growing tips are more frequent than in the valleys between tips because they have a lower effective-energy cost. During sprouting, conservation of cell area requires that the cells in the valleys must retract, while those in the tips protrude. In the Savill-Hogeweg algorithm, retraction is energetically favorable, while it is energetically neutral in our pseudopod-extension-only chemotaxis algorithm, making the net change in effective energy positive with a rate depending on the cell motility. The effective-energy change is negative in the Savill-Hogeweg algorithm and thus nearly independent of $T$ (Fig. 13 where $H_0$ is the initial effective energy).

# 3. Discussion

We have shown that a single set of cell behaviors, *i.e.* contact-inhibited chemotaxis to an autocrine, secreted chemoattractant can explain aspects of both *de novo* and sprouting blood-vessel growth. Our results suggest that branching in aggregates of chemotacting ECs could result from two separate effects of the same mechanism. For low cell motilities $T$, *i.e.* a low probability for active, dissipative cellular protrusion, the branching resembles a buckling instability (see *e.g.* [42]), in which the surface cells exert a surface-normal force on the cluster's inner core. For larger cell motilities, the shallower chemoattractant gradients at protrusions make the ECs there more likely to extend outward-directed pseudopods than cells in the valleys between the protrusions.

While we have adopted the Torino Group's assumption that ECs chemotax in response to gradients of a diffusible, autocrine, secreted chemoattractant [12,25], our simulation also reproduces continuum models that assume that ECs stress the ECM [15], which either pulls on the surrounding ECs, provides haptotactic cues for active EC migration [16], or both [26]. Because these models assume that ECs exert radially-symmetric stresses on the ECM, the underlying mathematical descriptions of the chemotactic and haptotactic mechanisms are equivalent. In both cases, contact inhibition should still operate and the patterning mechanism we have proposed should still apply, with traction or haptotaxis replacing chemotaxis and the mechanical screening length replacing the diffusion length. Our simulation may also apply to the formation of linear structures by non-vascular, glia or muscle cells cultured on rigid, plastic culture dishes in continuously-shaken medium [17] in which cells explore their environment using long filopodia, then move towards their neighbors by pulling themselves along bound filopodia. Thus, the combination of cell aggregation and contact-inhibition that drives patterning in our model, could also occur without chemical gradients and even without ECM.

Our simulations also allow us to clarify a number of subtleties concerning the interpretation of our own and others' experiments in which blocking VE-cadherin interfered with normal vascular

Contact-inhibited chemotaxis in *de novo* and sprouting blood vessel growth patterning. In our *in vitro* experiments, anti-VE-cadherin treatment caused ECs to round, in addition to its hypothesized effect on contact inhibition, so our experiments cannot rule out the possibility that the anti-VE-cadherin treatment inhibited vascular patterning because of its effect on EC shape. A further complication is that anti-VE-cadherin treatment could conceivably reduce the adhesion between ECs. As we noted above, In VE-cadherin -/- knock-out mice, ECs still form strong adhesive junctions [18], suggesting that VE-cadherin is not required for EC-EC binding.

Our simulations show that the contact-inhibition patterning mechanism operates over a wide range of cell-cell adhesions, suggesting that changes in adhesivity are not significant provided that contact-inhibition persists, and independent of cell shape[23], suggesting that the shape change is not significant. However, we have also shown that strong cell-cell adhesion plus chemotaxis can produce vascular-like patterns in simulations [21]. Fortunately, the three vascular patterning mechanisms (contact-inhibition, cell-elongation and cell-cell adhesion) have vastly different kinetics [22]. Thus time-lapse microscopy experiments [19,44] quantifying the kinetics of capillary-plexus development (see *e.g.* [22]), will allow us to definitively distinguish among these three patterning mechanisms. Already, we can say that adhesion-driven patterning is so slow and requires such strong adhesion that it appears incompatible with the available qualitative data from experiments.

To further test if VE-cadherin-mediated, contact-dependent signaling to VEGF-R2 [20], rather than VE-cadherin's function as a cell-adhesion molecule is responsible for the effects of anti-VE-cadherin treatment in mouse yolk sacs, we could experimentally block signal transduction from VE-cadherin to VEGFR-2, specifically interfering with VE-cadherin's signaling function, while leaving its role as an adhesion molecule intact. A possible target would be CD148, which phosphorylates VEGFR2 after VE-cadherin binding [20,45]. Embryonic vascularization and angiogenic sprouting are severely deficient in CD148 -/- knock-out mice [45], further supporting our hypothesis that VE-cadherin's contact-dependent intercellular signaling is crucial to vasculogenesis and angiogenesis.

Contact-inhibited chemotaxis in *de novo* and sprouting blood vessel growth

Perryn *et al*. [19] showed that anti-VE-cadherin treatment reduced sprout extension in murine allantois cultures by 70%, while it reduced cell-autonomous motility along sprout segments by 50 %. Based on these results, they postulated that VE-cadherin is required for the motility of ECs along sprouts towards the tip. However, our simulations show that the observed cell slow-down after anti-VE-cadherin administration may be an indirect effect of a reduction of sprouting. Furthermore, our simulations suggest that even substantially reduced cell motility may not prevent patterning, though it does slow it down.

In our simulations, branching and pattern formation require only experimentally-observed cell-level mechanisms, instead of the blood-vessel-level phenomenology in some other angiogenesis models [28-30]. However, by starting with a cluster of endothelial cells, our simulations ignore many events preceding sprout formation, including the release of plasma proteins by the vessel, the breakdown of the basal lamina, the detachment of the ECs from surrounding ECs and smooth muscle cells, and cell proliferation. They also ignore subsequent processes consolidating outgrowth of the sprout, including tip-cell selection, any long-range chemoattractants and chemorepellants that guide the vessel to its target, the formation of new basal lamina, the sprout's association with stabilizing cells including pericytes, lumen formation within the sprout, and flow-induced remodeling of the developed vasculature. The mechanism for sprouting and network formation we have proposed forms a firm basis for future, more complete models of angiogenesis which include basal lamina and pericytes. We are currently studying the formation of directed sprouts with proliferating ECs in response to additional chemoattractants or chemorepellants and analyzing the role of cell elongation during sprouting. We are also studying the effect of additional, cell-cell contact-dependent signaling mechanisms, including delta-notch tip-cell selection [3] and chemoattractant-response amplifying Eph receptor-ephrin ligand interactions [8].

# 4. Materials and Methods

## *4.1. The Glazier-Graner-Hogeweg (GGH) Model*

The GGH represents biological cells as patches of identical lattice indices $\sigma(\vec{x})$ on a square or triangular lattice, where each index uniquely identifies, or labels a single biological cell. Connections (*links*) between neighboring lattice sites of unlike index $\sigma(\vec{x}) \neq \sigma(\vec{x}')$ represent bonds between apposing cell membranes, where the *bond energy* is $J(\sigma(\vec{x}), \sigma(\vec{x}'))$, assuming that the types and numbers of adhesive cell-surface proteins determine $J$. A penalty increasing with the cell's deviation from a designated target volume $A_{\text{target}}(\sigma)$ imposes a *volume constraint* on the simulated ECs. We define the pattern's *effective energy*:

$$H = \sum_{neighbors} J(\sigma(\vec{x}), \sigma(\vec{x}'))(1 - \delta(\sigma(\vec{x}), \sigma(\vec{x}'))) + \lambda \sum_{\sigma} (a(\sigma) - A_{\text{target}}(\sigma))^2, \qquad (2)$$

where $\vec{x}$ and $\vec{x}'$ are neighboring lattice sites (up to fourth-order neighbors), $a$ is the current area of cell $\sigma$, $A_{\text{target}}(\sigma)$ is its target area, $\lambda$ represents a cell's resistance to compression, and the Kronecker delta is $\delta(x, y) = \{1, x = y; 0, x \neq y\}$. Each lattice site represents an area of $2\,\mu m \times 2\,\mu m$. Since we assume that ECs do not divide or grow during patterning, we set $A_{\text{target}}(\sigma) = 50$ lattice sites, corresponding to a cell diameter of about $16\,\mu m$, and $\lambda = 25$ for all cells. The ECs reside in a very thin layer of extracellular fluid, which is a generalized cell without a volume constraint and with $\sigma = 0$. We assume that the ECs and fluid sit on top of a rigid ECM through which the chemoattractant diffuses, but we do not represent this ECM in the GGH lattice. We also assume that the presence of the fluid does not disturb the chemoattractant distribution in the ECM. Unless we specify otherwise, we use a bond energy $J(c, c) = 40$ between the ECs, and $J(c, M) = 20$ between the ECs and the ECM. For these settings the ECs do not adhere without chemotaxis. We define a

Contact-inhibited chemotaxis in *de novo* and sprouting blood vessel growth special, high *cell-border energy* $J(c, B) = 100$ to prevent ECs from adhering to the lattice boundaries. We use fixed boundary conditions.

To mimic cytoskeletally-driven pseudopod extensions and retractions, we randomly choose a source lattice site $\vec{x}$, and attempt to copy its index $\sigma(\vec{x})$ into a randomly-chosen neighboring lattice site $\vec{x}'$. For better isotropy we select the source site from the twenty, first- to fourth-nearest neighbors [46]. During a *Monte Carlo Step* (*MCS*) we carry out $N$ *copy attempts*, with $N$ the number of sites in the lattice. We set the experimental time per MCS to $30\,\text{s}$; for this setting the simulated ECs move with nearly their experimental velocity [22]. We calculate how much the effective energy would change if we performed the copy, and accept the attempt with probability $P(\Delta H) = \{e^{\frac{-\Delta H}{T}}, \Delta H \geq 0; 1, \Delta H < 0\}$, where $T$ defines the *intrinsic cell motility*. All our simulations, except those in Figs. 11-13, use $T = 50$.

In experiments, cells respond to chemoattractant gradients by executing a more-or-less-strongly biased random walk up or down the gradient, where, over times short enough to allow us to neglect adaptation, the velocity of the drift depends on the gradient strength and the absolute concentration. We therefore define a set of extensions to the basic GGH model which reproduce these empirical behaviors due to preferential extension and retraction of pseudopods up chemoattractant gradients [24] by including a *chemical effective-energy change* at each copy attempt [21,36],

$$\Delta H_{\text{chemotaxis}} = -\mu \left( \frac{c(\vec{x}')}{1 + sc(\vec{x}')} - \frac{c(\vec{x})}{1 + sc(\vec{x})} \right), \quad (3)$$

where $c$ is the concentration of the chemoattractant, which we assume is present everywhere in a layer of ECM *under* the ECs, $\vec{x}'$ is the target site, $\vec{x}$ the source site, and $s$ regulates the saturation of the chemotactic response. Unless we specify otherwise, we set $s = 0$, in which case chemotaxis depends linearly on the chemoattractant gradient only, independent of the chemoattractant

Contact-inhibited chemotaxis in *de novo* and sprouting blood vessel growth concentration. $\Delta H_{\text{chemotaxis}} \rightarrow 0$ for large values of *s* and, for *s*≠0, for high chemical concentraycins. The *chemotaxis coefficient* is $\mu = \chi(c,M)$ at cell-ECM interfaces and $\mu = \chi(c,c)$ at cell-cell interfaces respectively. Setting $\chi(c,c) = 0$ and $\chi(c,M) = 500$ ensures that chemotactic extensions occur only at cell-ECM interfaces, reflecting VE-cadherin's suppression of pseudopods. Both extending and retracting pseudopods contribute to the chemical effective-energy change. To implement *pseudopod-extension-only chemotaxis* (see Figs. 11-13), where only *extending* pseudopods at the cell-ECM interface respond to the chemoattractant, cells experience a chemical effective-energy change only if the source lattice site $\vec{x}$ belongs to an EC, *i.e.*

$$\Delta H_{\text{chemotaxis}} = -(1 - \delta(\sigma(\vec{x}),0))\mu(\frac{c(\vec{x}')}{1 + sc(\vec{x}')} - \frac{c(\vec{x})}{1 + sc(\vec{x})}). \qquad (4).$$

For a more detailed discussion of chemotaxis in the GGH model see [47]. We solve the partial-differential equation for chemoattractant diffusion and degradation (Eq. 1) numerically using a finite-difference scheme on a lattice matching the GGH lattice. We use 15 diffusion steps per MCS, with $\Delta t = 2\,\text{s}$. For these parameters, the chemoattractant diffuses more rapidly than the ECs, enabling us to ignore advection in the medium as the cells push the fluid.

Source code for the simulations is available online from `http://sourceforge.net/projects/tst`. Parameter files for the simulations in this paper are included in the online supplementary material.

### *4.2. Allantois Culture and Immunolabeling*

We dissected allantoides from mouse embryos at embryonic stages 7.5-8.0. We washed the explants in fresh, cold ePBS and pipetted them into fibronectin-coated (5 mg/ml) Delta-T culture dishes (Bioptechs, Butler, PA) containing high-glucose, phenol-red-free Dulbecco's modified Eagles' medium (*DMEM*) supplemented with 10% fetal bovine serum (*FBS*), 1% penicillin-streptomycin, and 1% L-glutamine (GibcoBRL, Grand Island, NY). We maintained the allantoic explants using standard culture conditions ($37\,^{o}C$ and 5% $CO_2$ / 95% air atmosphere) in a custom-designed culture chamber for 12-24 hours in the presence of an endothelial-specific marker, CD34 monoclonal antibody (BD PharMingen, San Diego, CA) directly conjugated to Cy3 (Amersham Biosciences). We fixed the allantoides in 3% paraformaldehyde for 20 minutes at room temperature, followed by an ePBS wash. For VE-cadherin antibody perturbations, we added anti-VE-cadherin monoclonal antibody (BD PharMingen, San Diego, CA) at 25 μg / ml to the culture medium.

*4.3. Image Acquisition*

We observed the cultures with a $10\times$ objective (0.30 N.A.) on an inverted, automated, wide-field, epifluorescence/differential-interference-contrast (*DIC*) microscope (Leica DMIRE2, Leica Microsystems, Germany). We recorded images ($608\times512$ pixel spatial and 12-bit intensity resolution) with a cooled Retiga 1300 camera (QImaging, Burnaby, British Columbia) in $2\times2$ binned acquisition mode, using 100-300 ms exposures. Image acquisition and microscope settings used software described in [44].

# 5. Acknowledgments

We gratefully acknowledge discussions with Stuart A. Newman, András Czirók, and Charles D. Little, and helpful comments of the anonymous referees. The simulation software is based on codes RM developed during work with Paulien Hogeweg at Utrecht University, The Netherlands. This work received support from NIH grants NIGMS 1R01 GM077138-01A1 and 5R01 GM076692 (JAG; AS), the Indiana University Biocomplexity Institute (RM), the Indiana University AVIDD program (JAG; RM) and an American Heart Association predoctoral fellowship 0410084Z (EDP).

# 6. Supporting Information

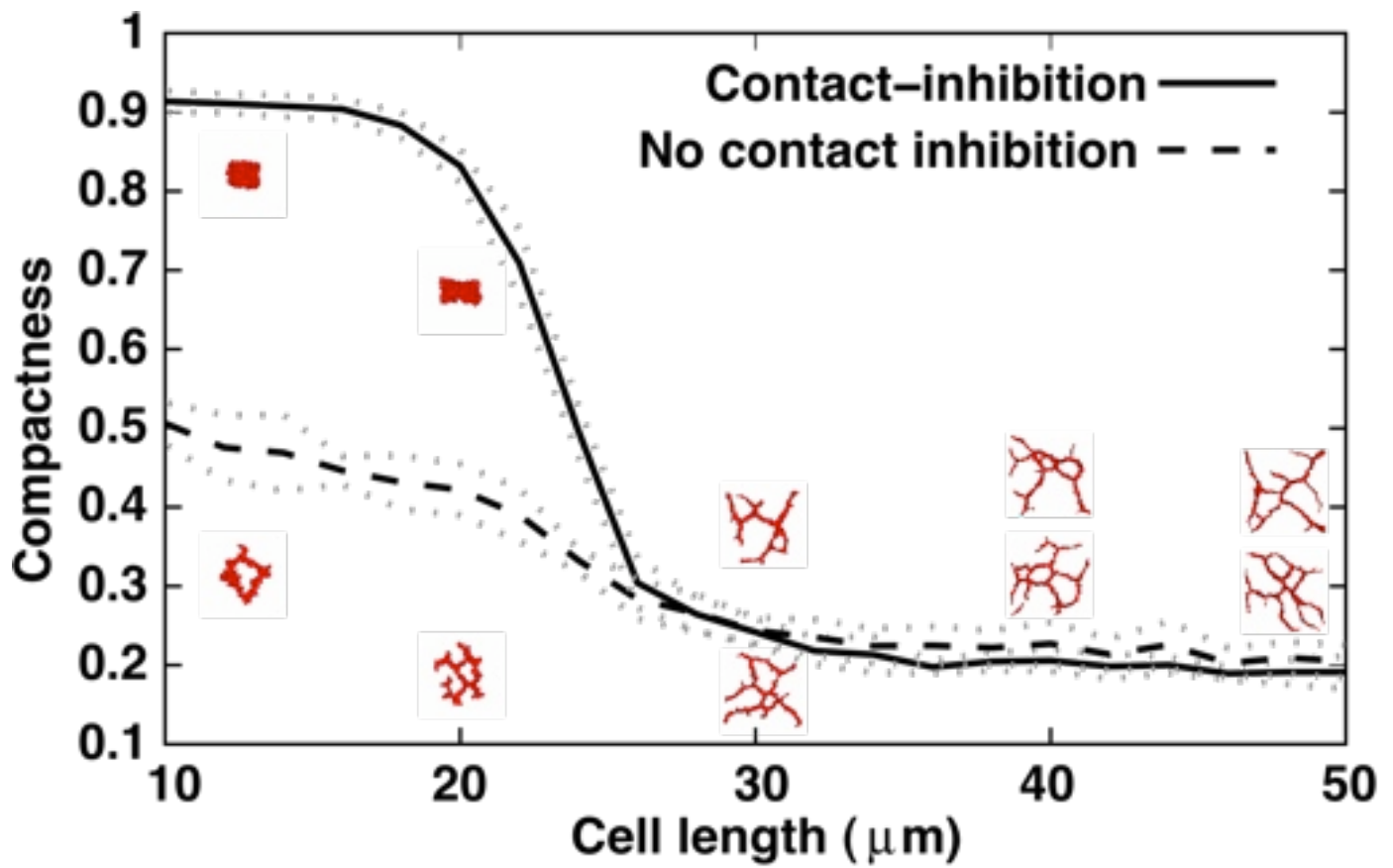

**Fig. S1.** Compactness ($C = A_{cluster}/A_{hull}$) of 128-cell clusters after 5000 MCS (~ 40 h) as a function of the cell length, in the presence (solid line) or absence (dashed line) of contact inhibition. Lengths given in terms of the *target length* $\Lambda$ as defined in Merks *et al.* 2006 [22]. Grey lines show standard deviations over ten simulations.

Contact-inhibited chemotaxis in *de novo* and sprouting blood vessel growth

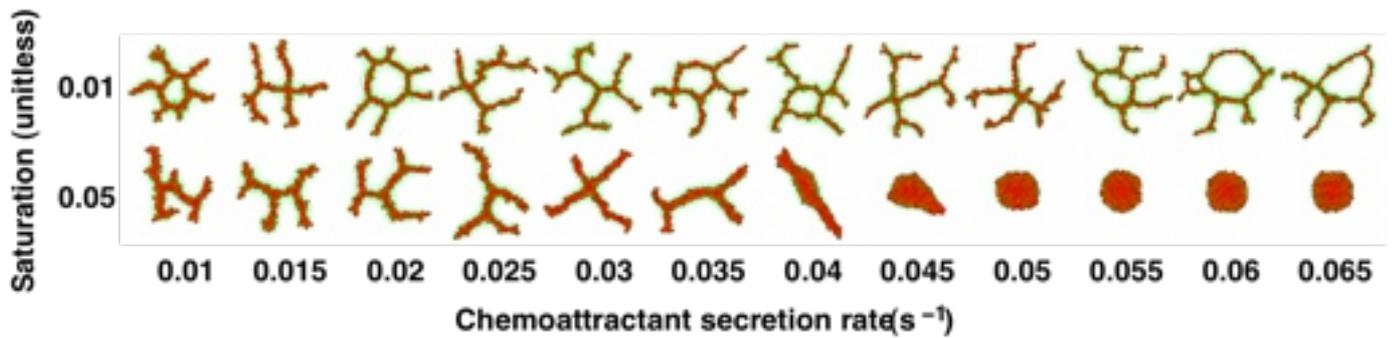

**Fig. S2.** Configurations of 128-cell clusters after 5000 MCS (~ 40 h) for increasing chemoattractant secretion rates for low (*s*=0.01) and high (*s*=0.05) chemoattractant receptor saturations.

*6.1 Supporting Movies are available online at*
*[http://www.psb.ugent.be/~romer/ploscompbiol](http://www.psb.ugent.be/~romer/ploscompbiol)*.

**Movie S1.** Endothelial cell aggregation without contact-inhibited chemotaxis. Simulation initiated with 1000 scattered cells. 0 MCS to 20,000 MCS (~ 0-170 h), 100 MCS per frame.

**Movie S2.** Endothelial cell aggregation in with contact-inhibited chemotaxis. Simulation initiated with 1000 scattered cells. 0 MCS to 2000 MCS (~ 0-20 h), 10 MCS per frame.

**Movie S3.** Same simulation as Movie S2. 0 MCS to 20,000 MCS (~ 0-170 h), 100 MCS per frame.

**Movie S4.** Sprouting instability in a simulation with contact-inhibited chemotaxis, initiated with a cluster of 256 endothelial cells. 0 MCS to 2000 MCS (~ 0-20 h), 10 MCS per frame.

**Movie S5.** Same simulation as Movie S4. 0 MCS to 20,000 MCS (~ 0-170 h), 100 MCS per frame.

**Movie S6.** Simulation with non-contact-inhibited chemotaxis, initiated with a cluster of 256 endothelial cells. 0 MCS to 2000 MCS (~ 0-20 h), 10 MCS per frame

Contact-inhibited chemotaxis in *de novo* and sprouting blood vessel growth

**Movies S7A-P.** Effect of cell adhesion on sprouting angiogenesis (*cf.* Fig. 7). Simulations with contact-inhibited chemotaxis, initiated with a cluster of 256 endothelial cells. 0 MCS to 20,000 MCS (~ 0-170 h), 100 MCS per frame. (A) *J(c,c)*=5; (B) *J(c,c)*=10; (C) *J(c,c)*=15; (D) *J(c,c)*=20; (E) *J(c,c)*=25; (F) *J(c,c)*=30; (G) *J(c,c)*=35; (H) *J(c,c)*=40; (I) *J(c,c)*=45; (J) *J(c,c)*=50; (K) *J(c,c)*=55; (L) *J(c,c)*=60; (M) *J(c,c)*=65; (N) *J(c,c)*=70; (O) *J(c,c)*=75; (P) *J(c,c)*=80.

**Movies S8A-K.** Effect of the chemotactic strength on sprouting angiogenesis (*cf.* Fig. 8). Simulations with contact-inhibited chemotaxis, initiated with a cluster of 256 endothelial cells. 0 MCS to 20,000 MCS (~ 0-170 h), 100 MCS per frame. (A) $\chi(c,M)=0$; (B) $\chi(c,M)=500$; (C) $\chi(c,M)=1000$; (D) $\chi(c,M)=1500$; (E) $\chi(c,M)=2000$; (F) $\chi(c,M)=2500$; (G) $\chi(c,M)=3000$; (H) $\chi(c,M)=3500$; (I) $\chi(c,M)=4000$; (J) $\chi(c,M)=4500$; (K) $\chi(c,M)=5000$.

**Movies S9A-F.** Effect of chemotaxis saturation on sprouting angiogenesis (*cf.* Fig. 9). Simulations with contact-inhibited chemotaxis, initiated with a cluster of 256 endothelial cells. 0 MCS to 20,000 MCS (~ 0-170 h), 100 MCS per frame. (A) *s*=0.0; (B) *s*=0.05; (C) *s*=0.1; (D) *s*=0.15; (E) *s*=0.2; (F) *s*=0.25.

**Movies S10A-G.** Effect of the diffusion constant *D* on sprouting angiogenesis (*cf.* Fig. 10). Simulations with contact-inhibited chemotaxis, initiated with a cluster of 256 endothelial cells. 0 MCS to 20,000 MCS (~ 0-170 h), 100 MCS per frame. (A) $D=1\cdot10^{-14}\,\text{m}^2\text{s}^{-1}$; (B) $D=5\cdot10^{-14}\,\text{m}^2\text{s}^{-1}$; (C) $D=1\cdot10^{-13}\,\text{m}^2\text{s}^{-1}$; (D) $D=2\cdot10^{-13}\,\text{m}^2\text{s}^{-1}$; (E) $D=3\cdot10^{-13}\,\text{m}^2\text{s}^{-1}$; (F) $D=4\cdot10^{-13}\,\text{m}^2\text{s}^{-1}$; (G) $D=5\cdot10^{-13}\,\text{m}^2\text{s}^{-1}$.

Roeland Merks *et al.* Page 29 30/05/08

Contact-inhibited chemotaxis in *de novo* and sprouting blood vessel growth

**Movies S11A-G.** Effect of the diffusion constant $D$ on the proposed sprouting-angiogenesis mechanism (*cf.* Fig. 10). Simulations with contact-inhibited chemotaxis, initiated with a cluster of 1024 endothelial cells. MCS 0 to 20,000 (~ 0-170 h), 100 MCS per frame. (A) $D = 1 \cdot 10^{-14} \, \text{m}^2 \text{s}^{-1}$; (B) $D = 5 \cdot 10^{-14} \, \text{m}^2 \text{s}^{-1}$; (C) $D = 1 \cdot 10^{-13} \, \text{m}^2 \text{s}^{-1}$; (D) $D = 2 \cdot 10^{-13} \, \text{m}^2 \text{s}^{-1}$; (E) $D = 3 \cdot 10^{-13} \, \text{m}^2 \text{s}^{-1}$; (F) $D = 4 \cdot 10^{-13} \, \text{m}^2 \text{s}^{-1}$; (G) $D = 5 \cdot 10^{-13} \, \text{m}^2 \text{s}^{-1}$.

**Movie S12.** Sprouting of a 256-cell cluster on a $500 \times 500$-pixel lattice (~$1 \, \text{mm} \times 1 \, \text{mm}$) with standard Savill-Hogeweg, extension-retraction chemotaxis at $T$=50, as in Fig. 12.

**Movie S13.** Sprouting of a 256-cell cluster on a $500 \times 500$-pixel lattice (~$1 \, \text{mm} \times 1 \, \text{mm}$) with standard Savill-Hogeweg, extension-retraction chemotaxis at $T$=200, as in Fig. 12.

**Movie S14.** Sprouting of a 256-cell cluster on a $500 \times 500$-pixel lattice (~$1 \, \text{mm} \times 1 \, \text{mm}$) with extension-only chemotaxis at $T$=200, as in Fig. 12.

**Protocol S1.** Tissue Simulation Toolkit v0.1.3. The source code for the software used for the simulations presented in this paper is also available from http://sourceforge.net/projects/tst. Installation: Unpack and compile according to the instructions given in the INSTALL file. The code is written in C++ using the cross-platform (Windows, Mac or Unix/Linux) library Qt (available from www.trolltech.com).

**Dataset S1.** Parameter files for the simulations shown in Figs. 1, 4, 11 and 12, packed as a tar.gz archive. To use, unpack the parameter-file archive and install the Tissue Simulation Toolkit (Protocol S1). Run the simulations from the command line using the command "vessel [parameter-file]". Reproduce the other simulations by editing the parameter files using a standard text editor to set the values specified in the text.

Contact-inhibited chemotaxis in *de novo* and sprouting blood vessel growth

Contact-inhibited chemotaxis in *de novo* and sprouting blood vessel growth

Contact-inhibited chemotaxis in *de novo* and sprouting blood vessel growth

**Figure legends**

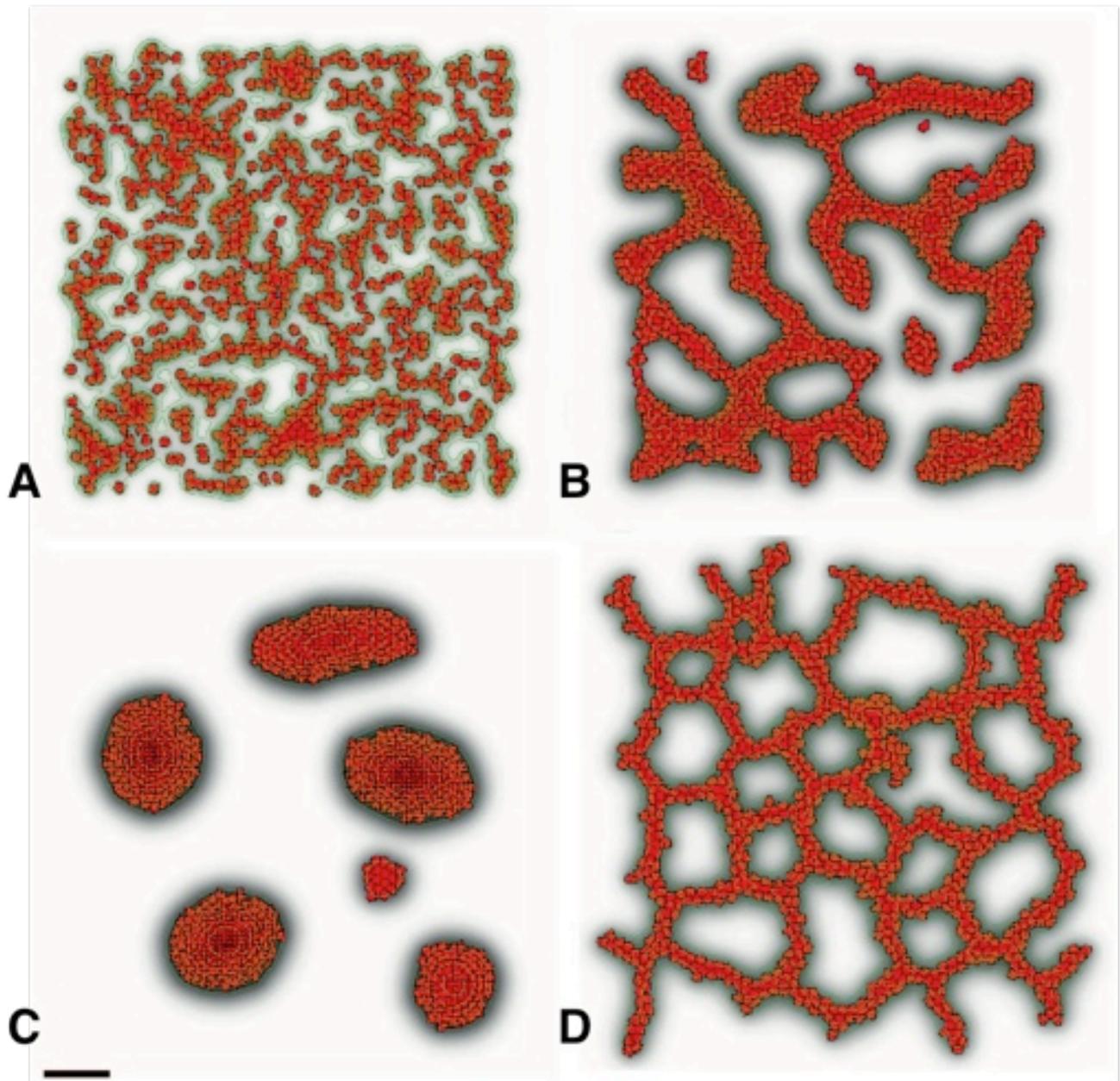

Fig. 1. Endothelial cell aggregation; simulation initiated with 1000 scattered cells: (A) After 10 Monte Carlo steps (*MCS*) (~ 5 min). (B) After 1000 MCS (~ 8 h). (C) After 10000 MCS (~ 80 h). (D) Contact-inhibited chemotaxis drives formation of vascular networks. Scale bar: 50 lattice sites ($\approx 100 \, \mu m$). Contour lines (green) indicate ten chemoattractant levels relative to the maximum concentration in the simulation. Grey shading indicates absolute concentration on a saturating scale.

Contact-inhibited chemotaxis in *de novo* and sprouting blood vessel growth

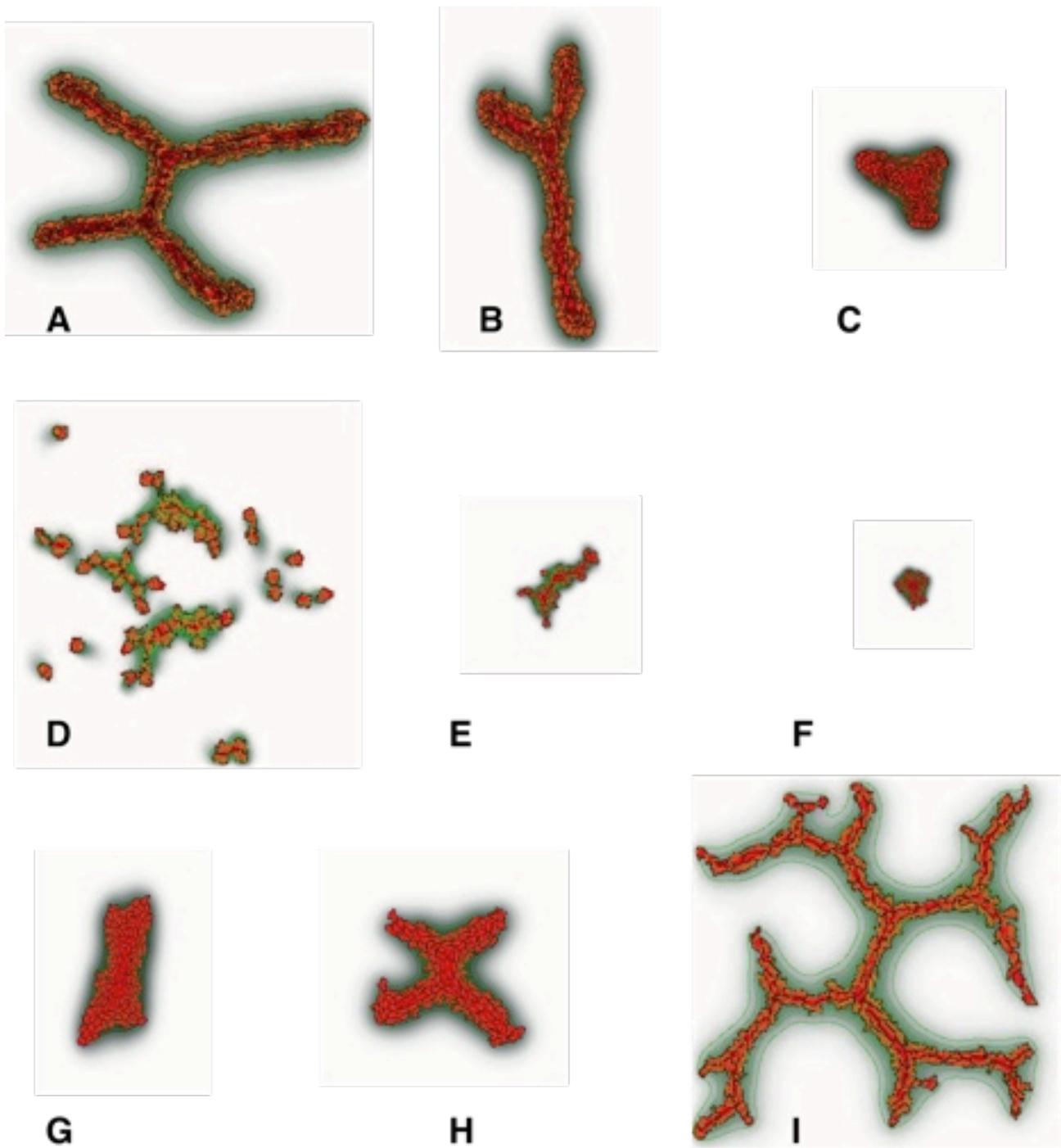

Fig. 2. Sprout formation in the absence of contact inhibition. (A-C) Adhesion-driven sprouting. (A) $J(c,c)=1$; (B) $J(c,c)=5$; (C) $J(c,c)=10$; (D-F) Passive cell elongation at short diffusion lengths; (D) $D=1\cdot 10^{-14}$ m$^2$s$^{-1}$; (E) $D=2\cdot 10^{-14}$ m$^2$s$^{-1}$; (F) $D=3\cdot 10^{-14}$ m$^2$s$^{-1}$; (G-I) Cell-autonomous cell elongation; (G) $\Lambda = 22$ μm; (H) $\Lambda = 24$ μm; (I) $\Lambda = 32$ μm.

Contact-inhibited chemotaxis in *de novo* and sprouting blood vessel growth

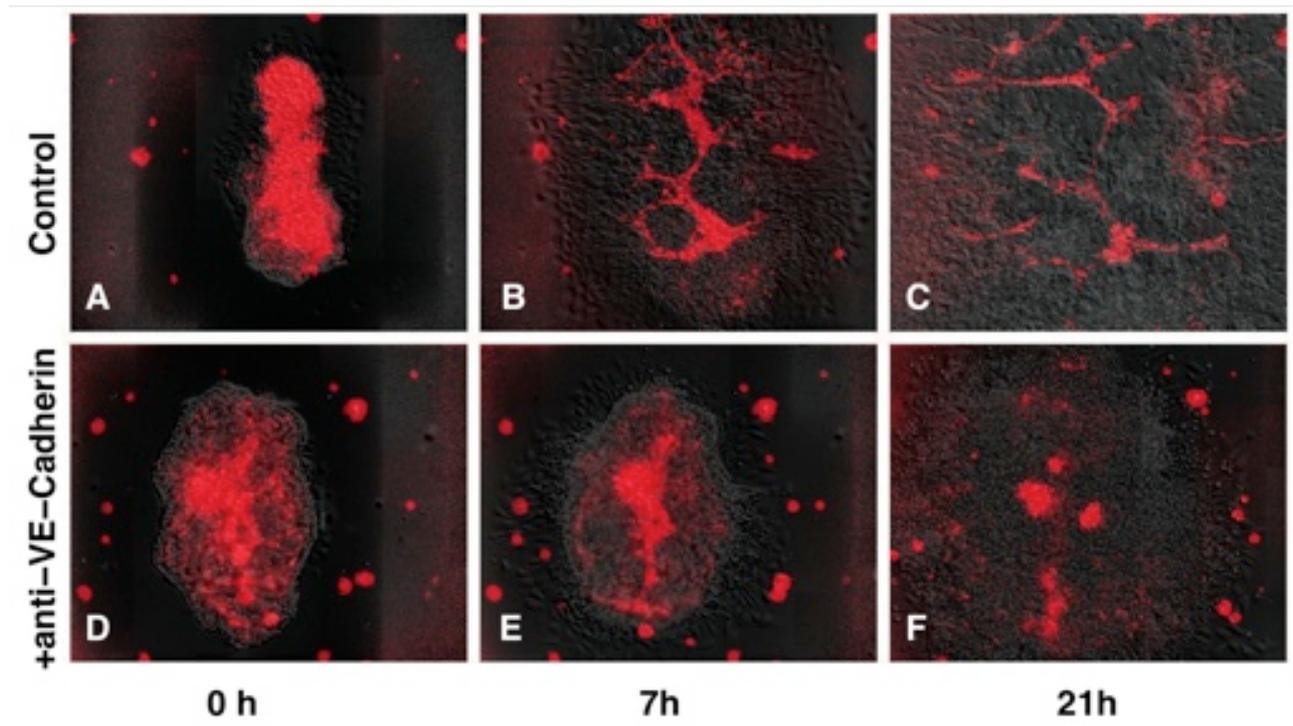

Fig. 3. Anti-VE-cadherin antibody treatment inhibits *de novo* blood-vessel growth in mouse allantois cultures. Endothelial cells fluorescently labeled in red with endothelium-specific CD34-Cy3 antibody. DIC/fluorescent image overlays. (A-C) Control. (D-F) Anti-VE-cadherin-treated cell cultures.

Contact-inhibited chemotaxis in *de novo* and sprouting blood vessel growth

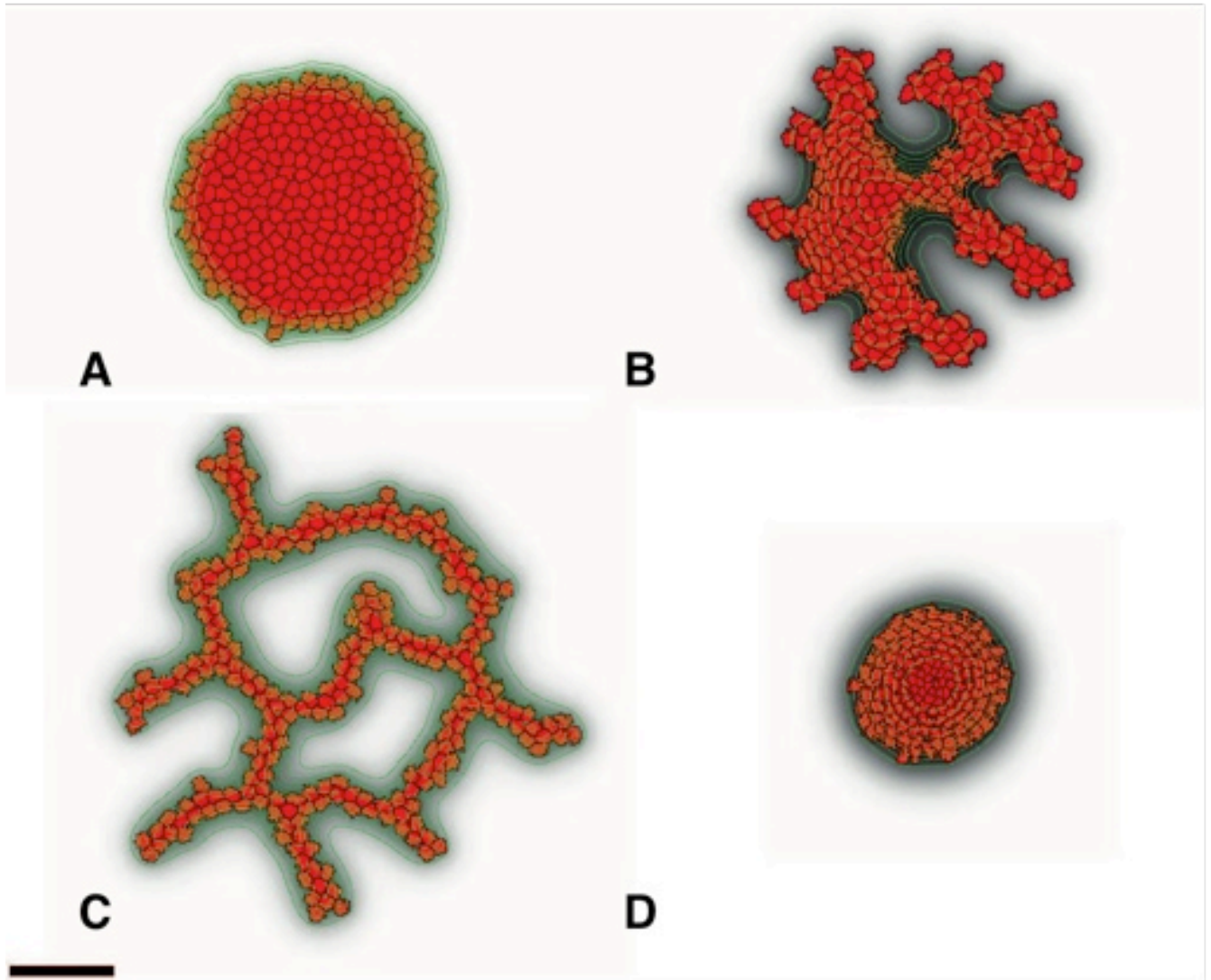

Fig. 4. Sprouting instability in a simulation initiated with a cluster of endothelial cells: (A) After 10 MCS (50 min). (B) After 1000 MCS (~ 8 h). (C) After 10,000 MCS (~ 80 h). (D) No sprouting in a simulation without contact inhibition of chemotaxis ($\chi(c,c)/\chi(c,M)=1$) at 10,000 MCS (~ 80 h). Scale bar: 50 lattice sites (~ 100 μm).

Contact-inhibited chemotaxis in *de novo* and sprouting blood vessel growth

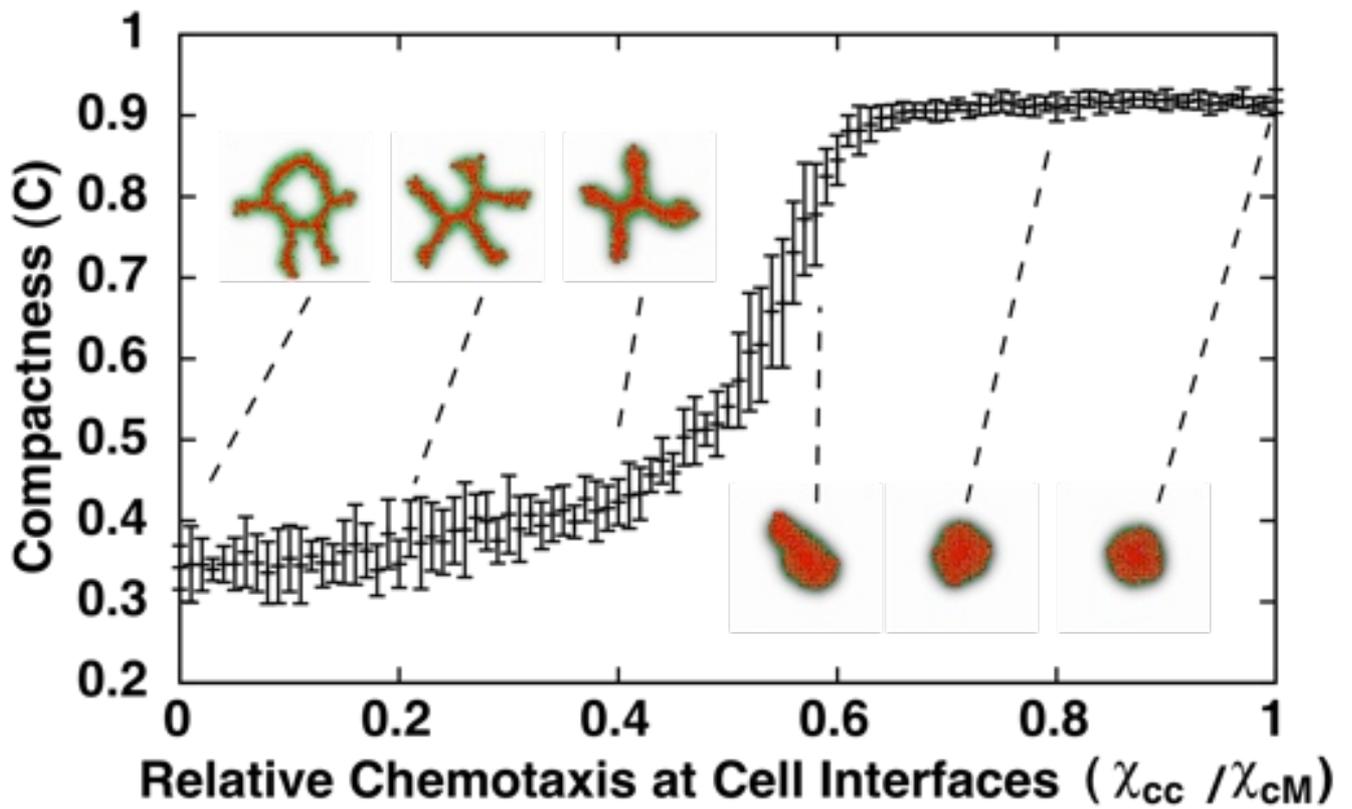

Fig. 5. Compactness ($C = A_{cluster}/A_{hull}$) of 128-cell clusters after 10,000 MCS (~80 h) as a function of the relative chemotactic response at cell-cell *vs.* cell-ECM interfaces. Error bars show standard deviations over ten simulations.

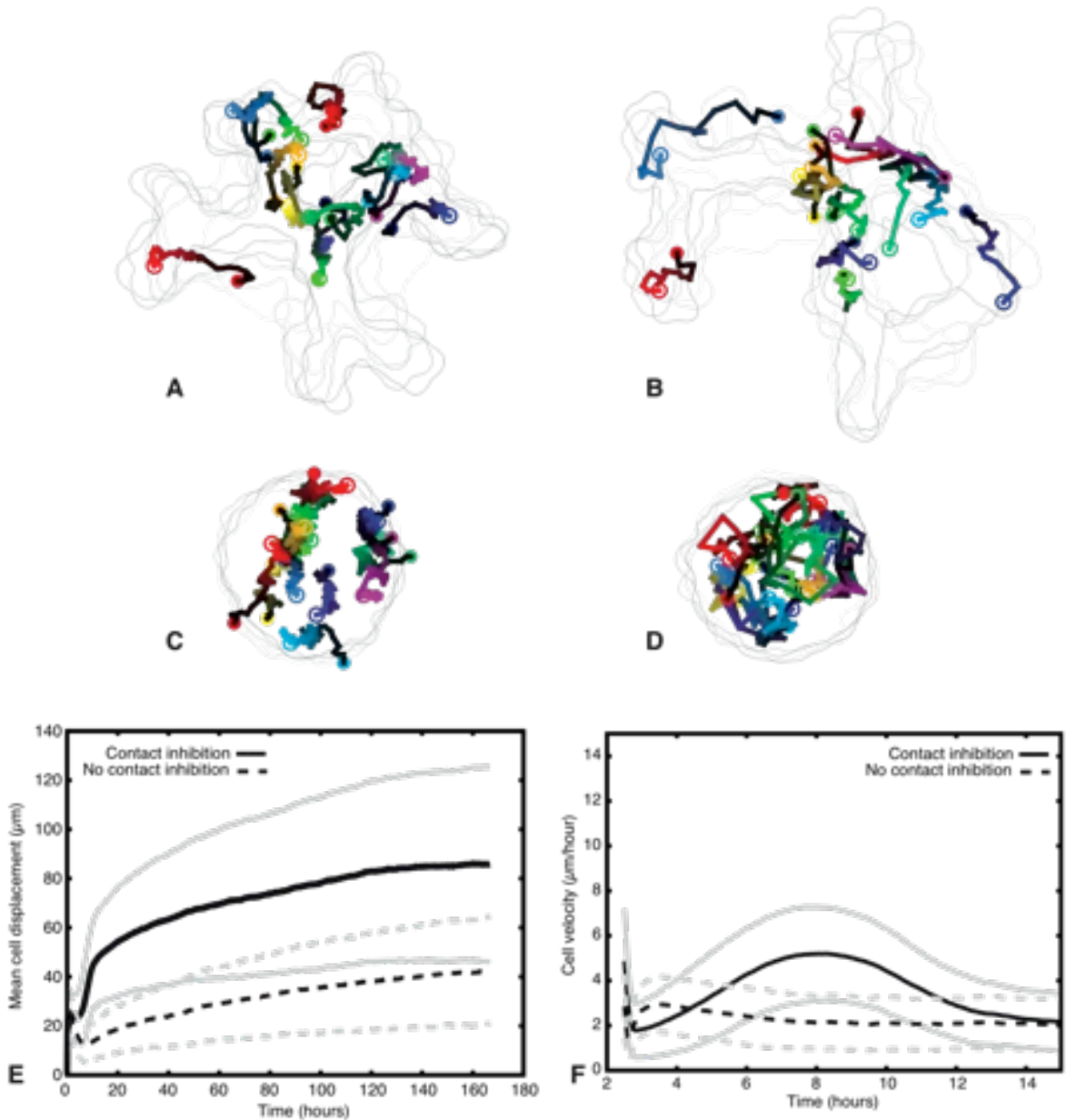

Fig. 6. Cell trajectories of simulated endothelial cells in 128-cell clusters in a contact-inhibited, sprouting cluster (A, B) and in a non-contact-inhibited, non-sprouting cluster (C, D). (A and C) Cell trajectories during initial sprouting, indicating the cells' centers of mass at 100 MCS (~ 50 min) intervals from 100 to 5000 MCS (~ 1-40 h). (B and D) Cell trajectories after initial sprouting, indicating the cells' centers of mass at 1000 MCS (~ 8 h) intervals from 4000 to 20,000 MCS (~ 30-170 h). Closed circles indicate initial cell positions; open circles indicate final cell positions. Colors identify individual cells; brightness increases from dark (initial positions) to bright (final positions). Outlines of clusters shown at 1000 MCS (~ 8 h) intervals (a,c) or 4000 MCS (~ 33 h) intervals

Contact-inhibited chemotaxis in *de novo* and sprouting blood vessel growth

(b,d). (e) Average displacement of cells from original positions over time in 10 simulations with 128 cells each, in contact-inhibited (solid curves) and non-contact-inhibited simulations (dashed curves). Grey lines indicate standard deviations. (f) Cell velocity $V_i(t) = (\vec{x}_i(t + \Delta t) - \vec{x}_i(t - \Delta t))/2\Delta t$ [19] with $\Delta t = 300$ MCS (~ 2.5 h) for contact-inhibited (solid curves) and non-contact-inhibited (dashed curves) simulations.

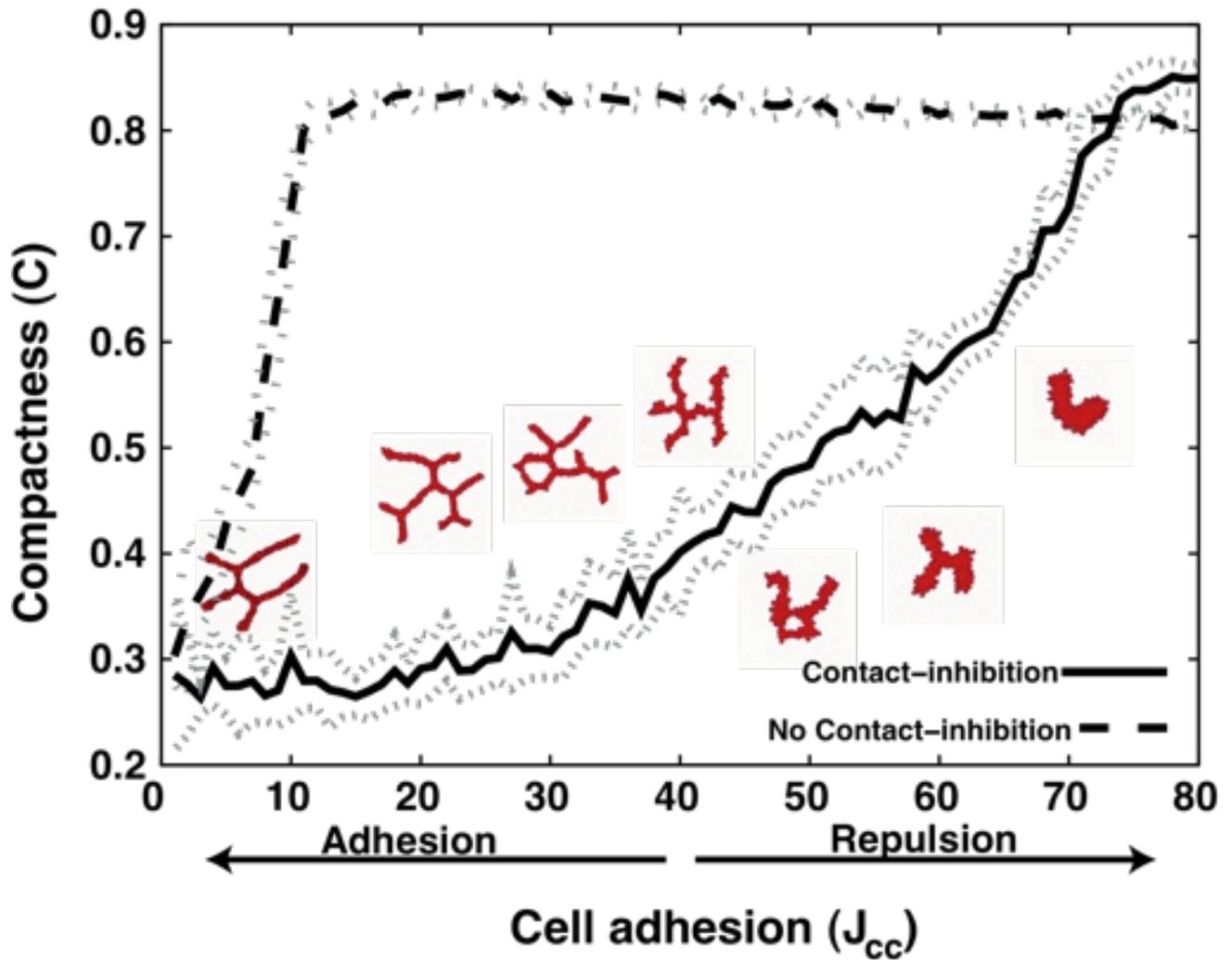

Fig. 7. Compactness ($C = A_{cluster}/A_{hull}$) of 128-cell clusters on $200 \times 200$-pixel lattices (~ $400\,\mu m \times 400\,\mu m$) after 5000 MCS (~ 40 h) for standard chemotaxis, as a function of the adhesion between endothelial cells, $J(c,c)$. For $J(c,c) < 40$ (i.e. $J(c,c) < 2J(c,M)$) the cells adhere without chemotaxis. Insets: Representative configurations after 5000 MCS (~ 40 h).

Contact-inhibited chemotaxis in *de novo* and sprouting blood vessel growth

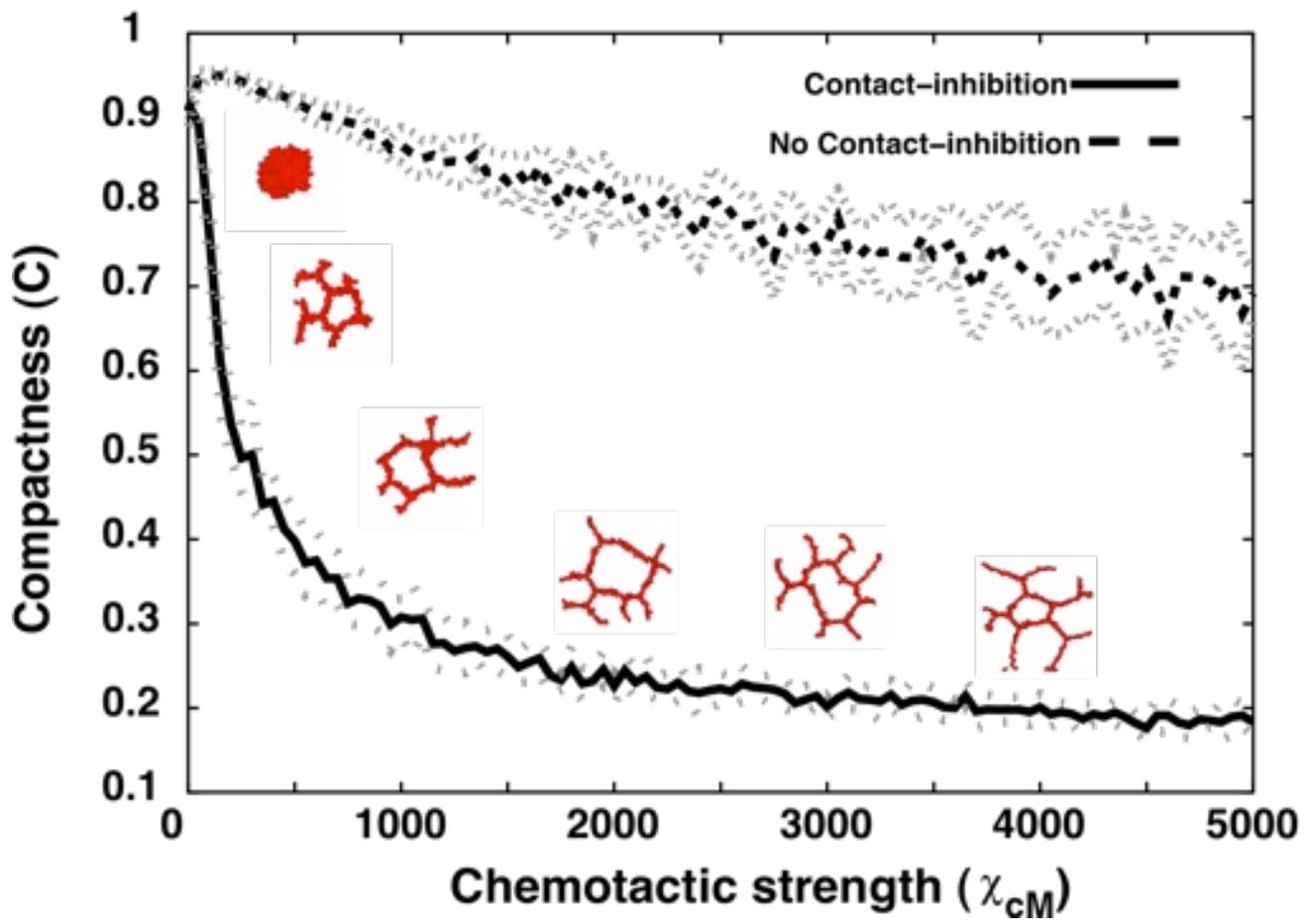

Fig. 8. Compactness ($C = A_{\text{cluster}}/A_{\text{hull}}$) of 128-cell clusters on $200 \times 200$-pixel lattices (~ $400\,\mu\text{m} \times 400\,\mu\text{m}$) after 5000 MCS (~ 40 h) for standard chemotaxis as a function of absolute chemotactic strength, $\chi(c,M)$. Insets: Representative configurations after 5000 MCS (~ 40 h).

Contact-inhibited chemotaxis in *de novo* and sprouting blood vessel growth

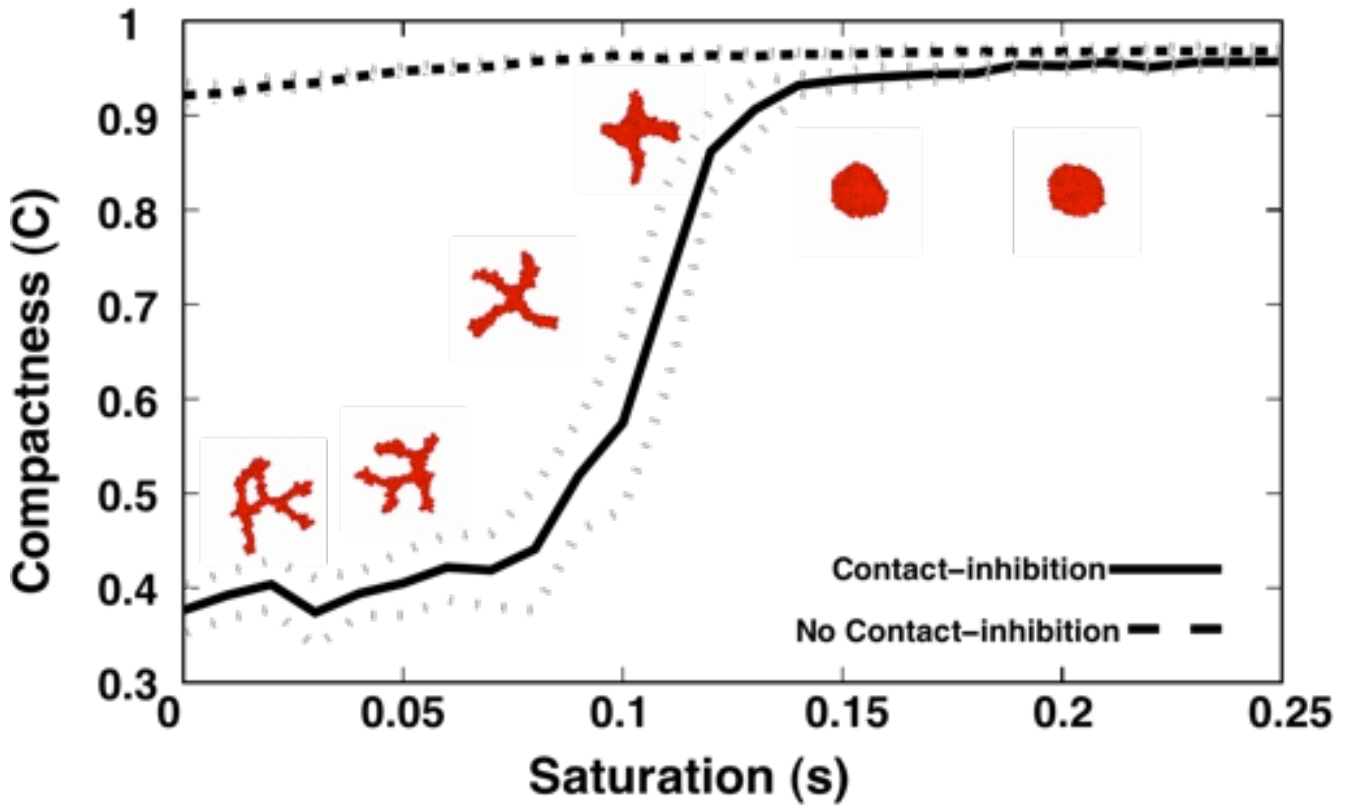

Fig. 9. Compactness ($C = A_{cluster}/A_{hull}$) of 128-cell clusters on $200 \times 200$-lattices (~ $400\,\mu m \times 400\,\mu m$) after 5000 MCS (~ 40 h) for standard chemotaxis as a function of the saturation of the chemotactic response, *s*. Insets: representative configurations after 5000 MCS (~ 40 h).

Contact-inhibited chemotaxis in *de novo* and sprouting blood vessel growth

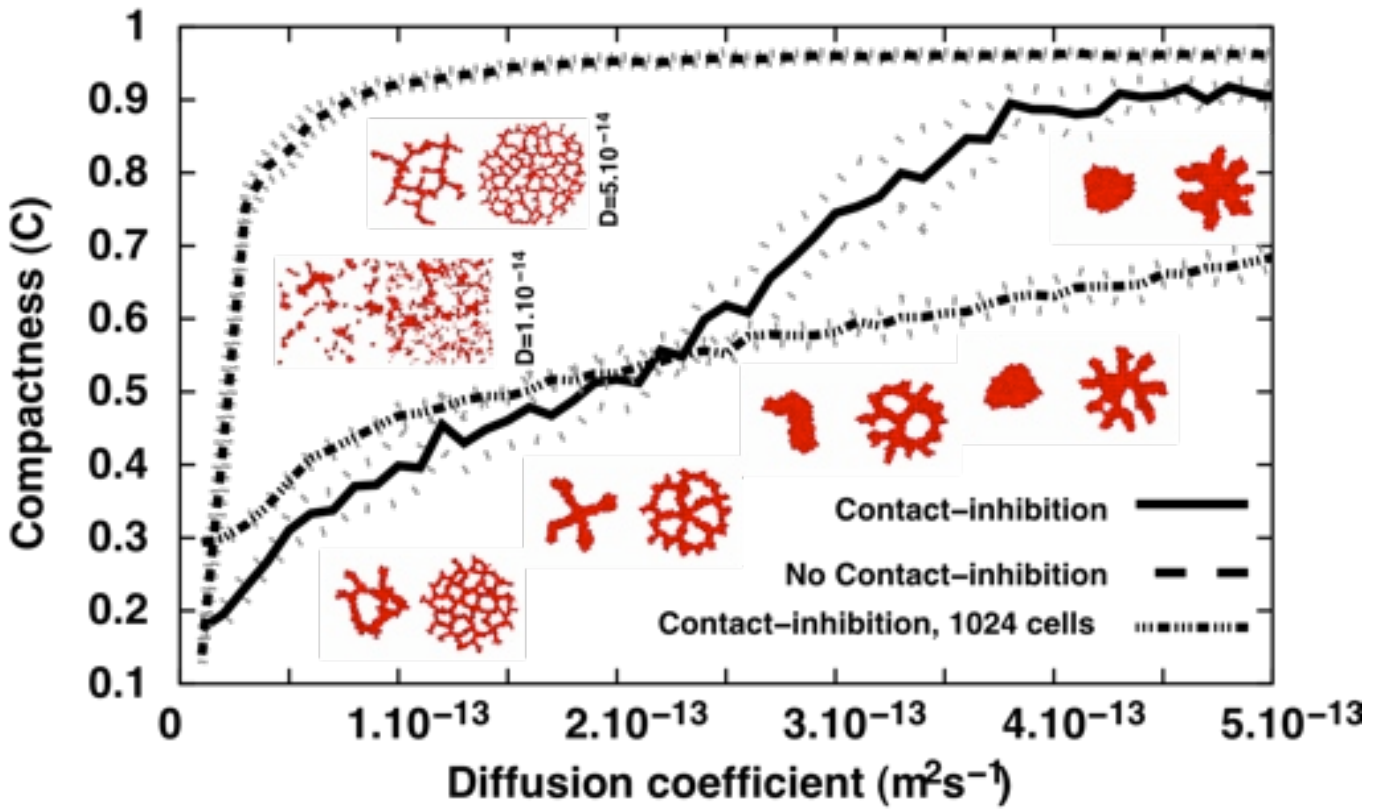

Fig. 10. Compactness ($C = A_{\text{cluster}}/A_{\text{hull}}$) of 128-cell clusters (solid curve) on $200 \times 200$-pixel lattices (~ $400\,\mu m \times 400\,\mu m$) and 1024-cell clusters (dashed curve) on $400 \times 400$-pixel lattices (~ $800\,\mu m \times 800\,\mu m$) after 5000 MCS (~ 40 h) for Savill-Hogeweg chemotaxis as a function of the chemoattractant diffusion constant *D*. Larger diffusion constants have longer diffusion lengths, $L = \sqrt{\dfrac{D}{\varepsilon}}$. Dashed-dotted line shows the compactness of VE-cadherin-inhibited 128-cell clusters. Insets: Representative configurations after 5000 MCS (~ 40 h) of the 128-cell clusters (left panels) and 1024-cell clusters (right panels; not to scale).

Contact-inhibited chemotaxis in *de novo* and sprouting blood vessel growth

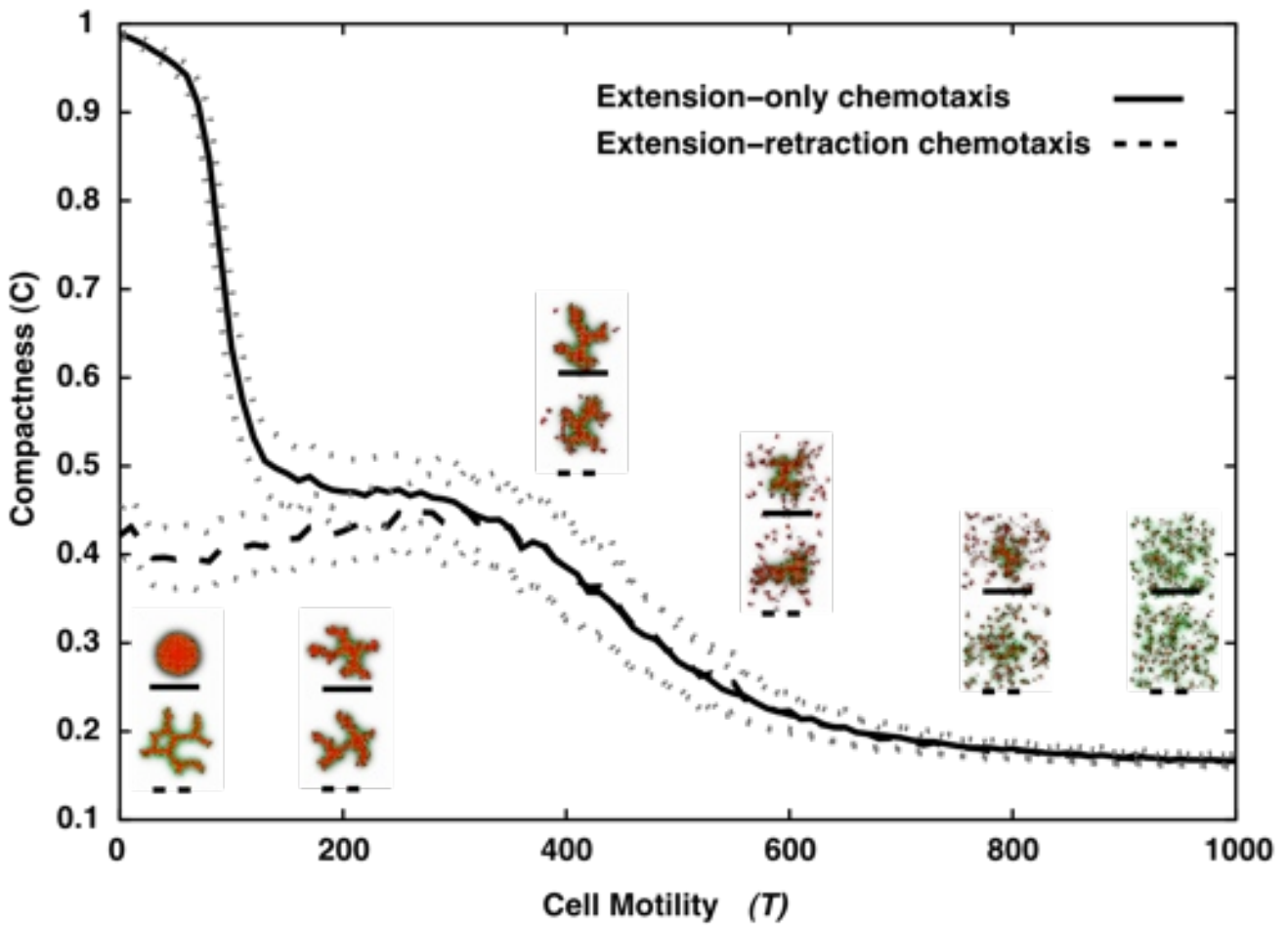

Fig. 11. Compactness ($C = A_{cluster}/A_{hull}$) of 128-cell clusters on $200 \times 200$-pixel lattices (~ $400\,\mu m \times 400\,\mu m$) after 5000 MCS (~40 h) as a function of the cell motility *T*, for standard Savill-Hogeweg [36] *extension-retraction* chemotaxis (solid line), and for *extension-only* chemotaxis (dashed line). Black lines show the mean over 100 simulations for each *T* (with a *T*-increment of 10). Dotted grey lines indicate one standard deviation. Insets: Representative configurations after 5000 MCS (~ 40 h).

Contact-inhibited chemotaxis in *de novo* and sprouting blood vessel growth

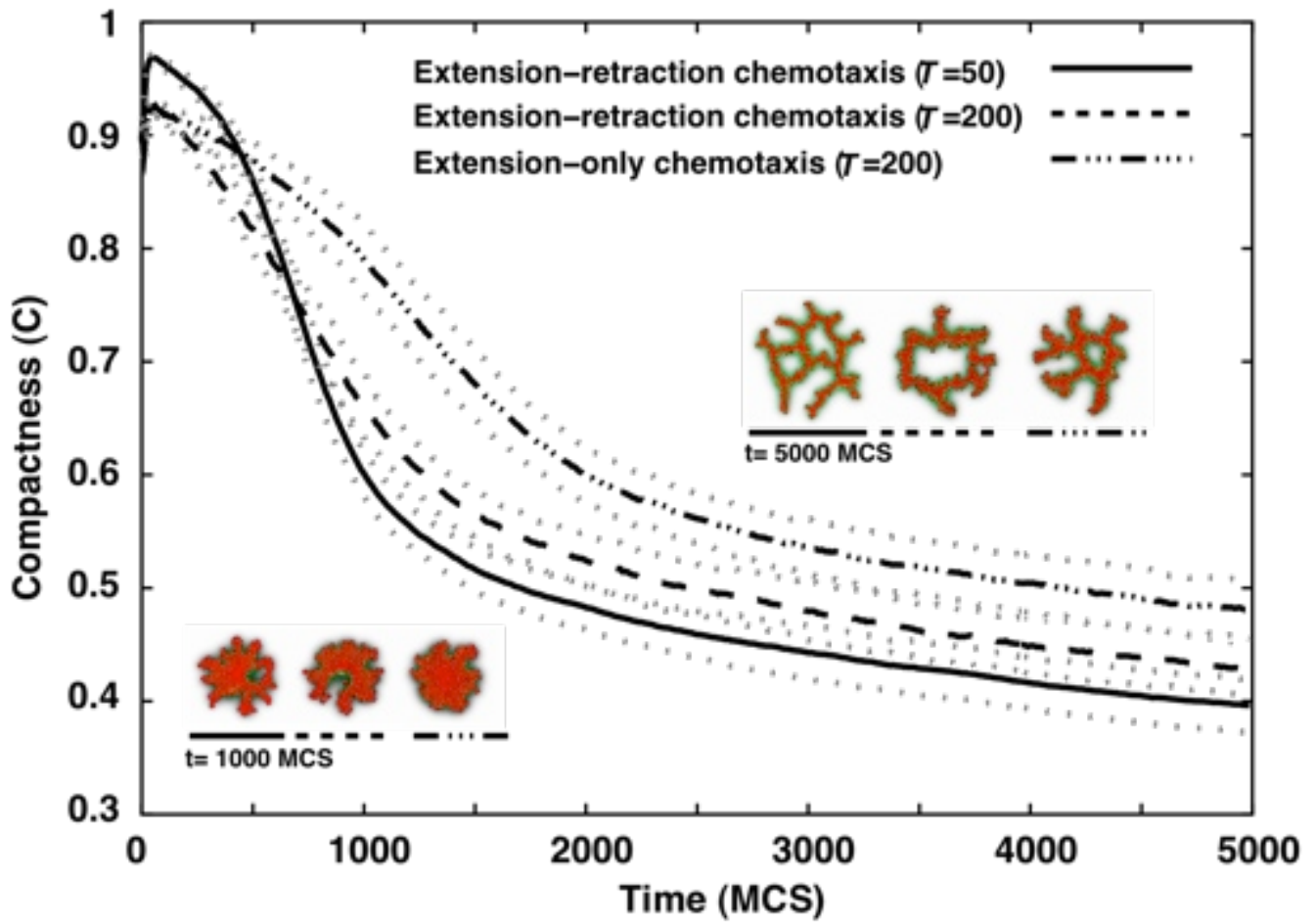

Fig. 12. Evolution of the compactness ($C = A_{\text{cluster}}/A_{\text{hull}}$) of 256-cell clusters on $500 \times 500$-pixel lattices (~ $1000\,\mu\text{m} \times 1000\,\mu\text{m}$) *vs.* time for standard Savill-Hogeweg [36] *extension-retraction* chemotaxis (solid and dashed lines, for $T = 50$ and $T = 200$ respectively), and for *extension-only* chemotaxis (dash-dotted line, $T = 200$), with only extending pseudopods responding to the chemoattractant. Black lines show the mean of 100 simulations. Dotted grey lines mark one standard deviation. Insets: Representative configurations after 1000 (~ 8 h) and 5000 MCS (~ 40 h). Movies available online.

Contact-inhibited chemotaxis in *de novo* and sprouting blood vessel growth

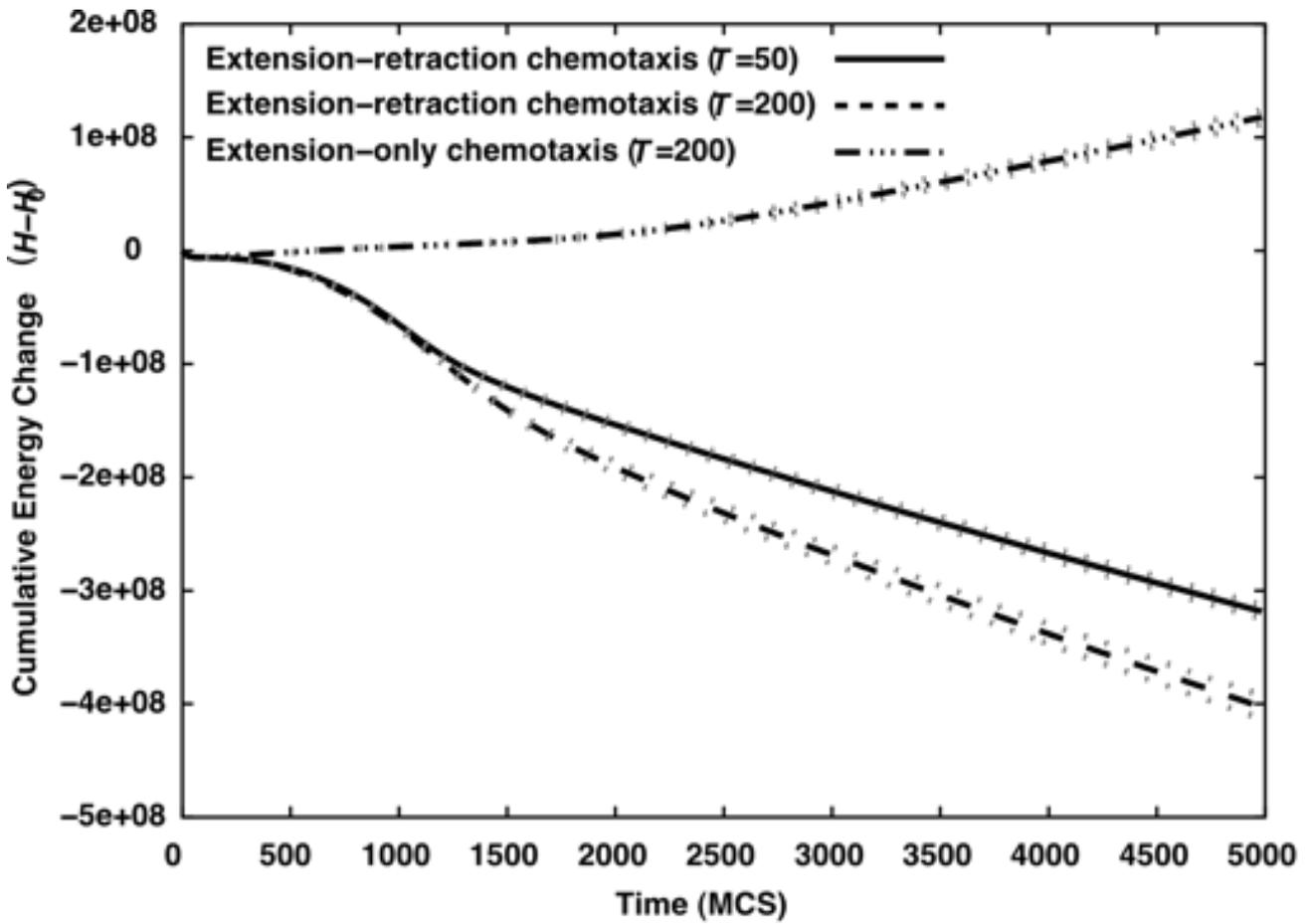

Fig. 13. Cumulative energy differences for standard Savill-Hogeweg [36] *extension-retraction* chemotaxis (solid and dashed lines, for $T = 50$ and $T = 200$ respectively), and for *extension-only* chemotaxis (dash-dotted line, $T = 200$) as a function of time. Black lines show the mean of 100 simulations. Dotted grey lines mark one standard deviation.